\documentclass[10pt,aps,pra,twocolumn,showpacs,superscriptaddress,floatfix]{revtex4-1}  % for review and submission
\usepackage{graphicx}  % needed for figureb
\usepackage{dcolumn}   % needed for some tables
\usepackage{bm}        % for math
\usepackage{amsmath}
\usepackage{amssymb}   % for math
\usepackage[utf8]{inputenc}
\usepackage{xfrac}
\usepackage{hyperref}

\newcommand{\ve}{\varepsilon}
\newcommand{\ket}[1]{\left|#1\right\rangle}

\newcommand{\create}[2]{{#1}^{\dagger}_{#2}}
\newcommand{\annihilate}[2]{{#1}^{\phantom{\dagger}}_{#2}}

\newcommand{\id}[1]{\mathrm{d}{#1}\,}
\newcommand\vvdots{\vphantom{\int^0}\smash[t]{\vdots}}

\usepackage{color}

\begin{document}

\title{Few photon scattering on Bose-Hubbard lattices}

\author{Kim G.~L.~Pedersen}
\author{Mikhail~Pletyukhov}
\email{pletmikh@physik.rwth-aachen.de}
\affiliation{Institut f\"{u}r Theorie der Statistischen Physik, RWTH Aachen University}

\date{\today}
 
\begin{abstract}
We theoretically investigate the scattering of few photon light on Bose-Hubbard lattices using diagrammatic scattering theory. We explicitly derive general analytical expressions for the lowest order photonic correlation functions, which we apply numerically to several different lattices. We focus specifically on non-linear effects visible in the intensity-intensity correlation function and explain bunching and anti-bunching effects in dimers, chains, rings and planes.

The numerical implementation can be applied to arbitrary Bose-Hubbard graphs, and we provide it as an attachment to this publication. 
\end{abstract}


\maketitle

\section{Introduction} % (fold)
\label{sec:introduction}
Experimentalists have in the last decade realized ever more complex quantum many-body photonic systems. This progress depends on several innovations, like the controlled guiding of photons in one-dimensional channels, their subsequent coupling to cavities, the coupling of multiple such cavities~\cite{Houck2012,Hoi2013,Hafezi2013a,Plotnik2014,Christodoulides2003}, and the creation of an effective interaction between photons through their mutual coupling to discrete quantum systems~\cite{Chang2007,Hafezi2012,VanLoo2013} or other non-linear media~\cite{Gupta2007}.

These experiments have attracted a lot of attention because of the possibility of realizing complex many-body quantum states~\cite{Angelakis2007,Hartmann2006,Greentree2006,Tomadin2010,Noh2017}; but this goal also has a backside, since the (exponentially) complex quantum many-body problem makes it increasingly difficult to perform meaningful theoretical calculations or exact numerical simulations.
Theory and experiment can, however, be brought on equal grounds by reducing the complexity of the involved quantum many-body states, as realized in the limit of weak input power, where only a few photons inhabit the scatterer at a time.

\begin{figure}[htb]
  \centering
  \includegraphics[width=.99\columnwidth]{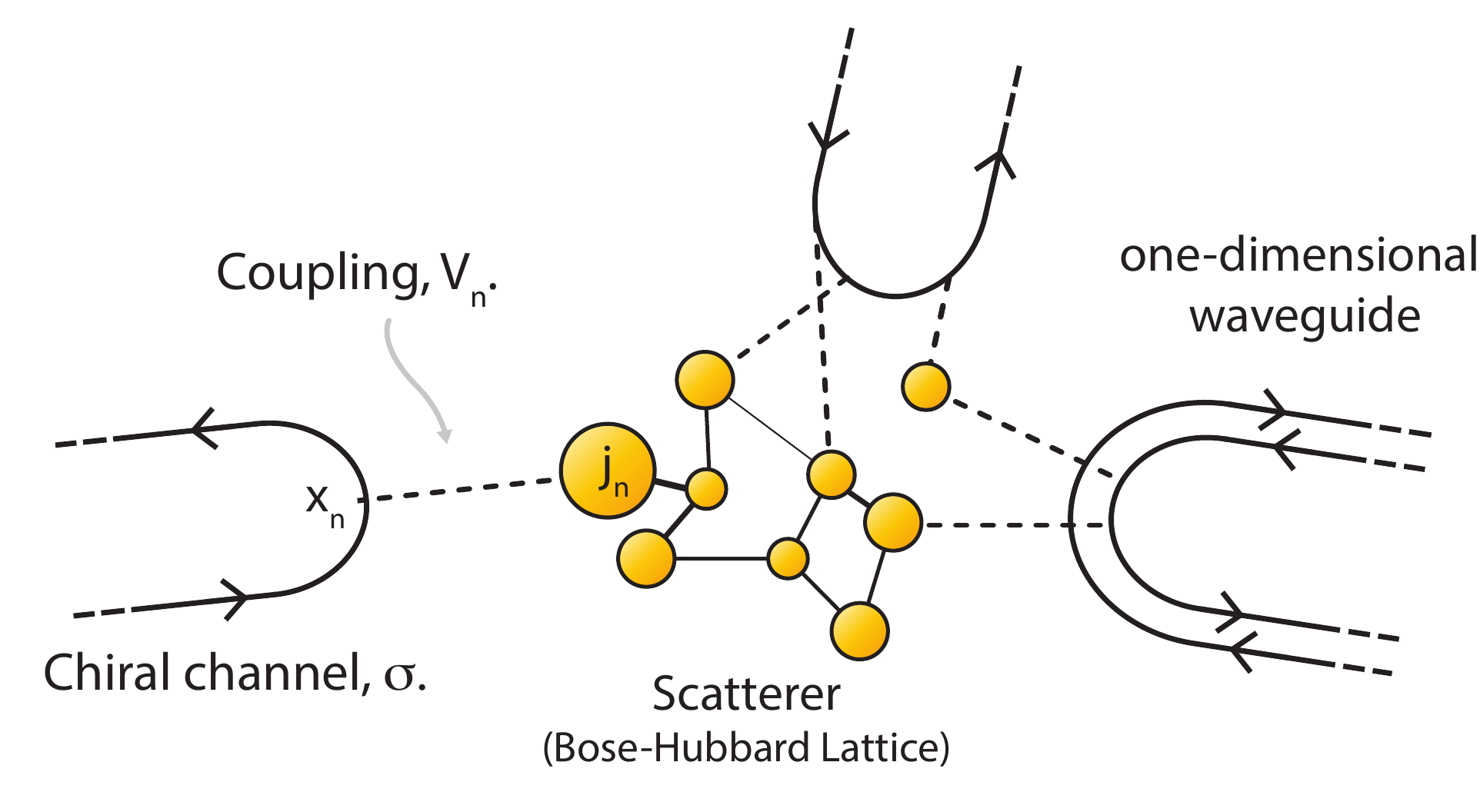}
  \caption{Example of a scattering setup: Four chiral channels (two of them forming a one-dimensional waveguide) couple at multiple points to an extended scatterer consisting of multiple sites.}
  \label{fig:model}
\end{figure}

We consider complex photonic systems composed of a distributed scatterer coupled to one or more one-dimensional chiral channels and operated in the few photon limit. Figure~\ref{fig:model} shows a general example of such a system, where the distributed scatterer is represented by connected dots, the chiral channels are symbolized by directed lines, while dashed lines indicate the couplings between the scatterer and channels.

The theory developed to describe such systems in the few photon limit exists in various incarnations as e.g. scattering theory~\cite{Pletyukhov2012a,Shen2007,Shen2007a,Lee2014,Fang2014}, Lehmann-Symanzik-Zimmermann formalism~\cite{Shi2009}, generalized master equations~\cite{Shi2015,Caneva2015}, SLH formalism~\cite{Brod2016,Brod2016a}, and input-output theory~\cite{Fang2014,Xu2015}. We here provide a generalization of the scattering theory methods to more complex local and quasi-local scatterers. While we here stop short of explicitly going beyond the Markov approximation~\cite{Zheng2013,Laakso2014,Baranger2015}, our results can be used to numerically solve a large class of photonic scattering problems, and used to understand and interpret many different photonic experiments operated in the weak power limit.

We here perform explicit numerical calculations for several different setups modeled as Bose-Hubbard graphs. While some of the distributed scatterer geometries have received some attention~\cite{Fang2014,Lee2014,Fang2016,Roy2013,Liew2010}, we also show how our method can be applied to larger systems such as chains, rings and square lattices. The paper is structured as follows:

Section~\ref{sec:model} describes the relevant theoretical models and discuss some of their possible experimental realizations.

Section~\ref{sec:diagrammatic_scattering_theory} reviews diagrammatic scattering theory within the Markov approximation. The diagrammatics are applied directly to the scattering  of few photons states on distributed quantum systems coupled locally and quasi-locally to the chiral channels. We explicitly derive the one-photon and two-photon scattering matrices, and finally find expressions for the first and second order correlation functions for coherent photonic input states. 

% Expressions for the coherence functions are provided in an appendix. 

Section~\ref{sec:results} investigates few photon scattering in several different Bose-Hubbard graphs using a numerical implementation of the results of the previous sections called {\tt babusca}\footnote{{\tt babusca} is also available at \href{http://github.com/georglind/babusca}.} and available in the supplementary information. For some simple systems we show that we easily reproduce the well-known results, and we then turn to more complicated geometries including dimers, chains and rings. We analyze the results and explain the main effects visible in the first and second order coherence functions, such as photon blockade, photon bunching and anti-bunching, and discuss their relationship to the geometry of the scatterer. We end the paper by demonstrating the capability of the {\tt babusca} approach by considering scattering an 8 by 8 square lattice scatterer.

% section introduction (end)
\section{Model} % (fold)
\label{sec:model}
We describe the relevant photonic systems by the general Hamiltonian,
\begin{align}
    H &= H_0 + V = H_{\text{sc}} + H_{\text{chs}} + V.
\end{align}
Here $H_{\text{sc}}$ describes the scatterer, which in our discussion is explicitly represented by a Bose-Hubbard lattice graph. $H_{\text{chs}}$ is a free Hamiltonian describing all the chiral one-dimensional channels, and $V$ describes the coupling between the channels and the scatterer.

Such photonic systems can be built in various experimental platforms, but one of the most direct implementations uses coupled cavity arrays based on either optical~\cite{Hafezi2013a} or superconducting microwave resonators~\cite{Houck2012,Hoi2013}, which in the presence of local non-linearities directly realize a Bose-Hubbard lattice model, as is also the case for multiple coupled qubits or emitters~\cite{Fink2009}. Other emitters - like atoms with intricate level structures - can also be modeled as effective Bose-Hubbard graphs, provided they support only a single ground state. 

The Bose-Hubbard lattice Hamiltonian can be written as the sum of three terms,
\begin{align}
  H_{\text{sc}} &= H_\ve + H_t + H_U \nonumber \\
   &= \sum_i \ve_i \create{b}{i} \annihilate{b}{i} + \sum_{i > j} (t_{ij} \create{b}{i} \annihilate{b}{j} + h.c.) \nonumber \\ 
   & \quad + \sum_i \tfrac{1}{2} U_{i} \create{b}{i} \annihilate{b}{i} (\create{b}{i} \annihilate{b}{i} -1).
\end{align}
Here the second quantization operators $\create{b}{i}$ ($\annihilate{b}{i}$) create (annihilate) an excitation on the lattice site $i$. Each lattice site has a local excitation energy, $\ve_i$, an onsite photon-photon interaction, $U_i$, also referred to as a nonlinearity; while the hopping between two neighbor lattices sites is captured by the amplitudes, $t_{ij}$.

\begin{figure}[b]
  \centering
  \includegraphics[width=0.8\columnwidth]{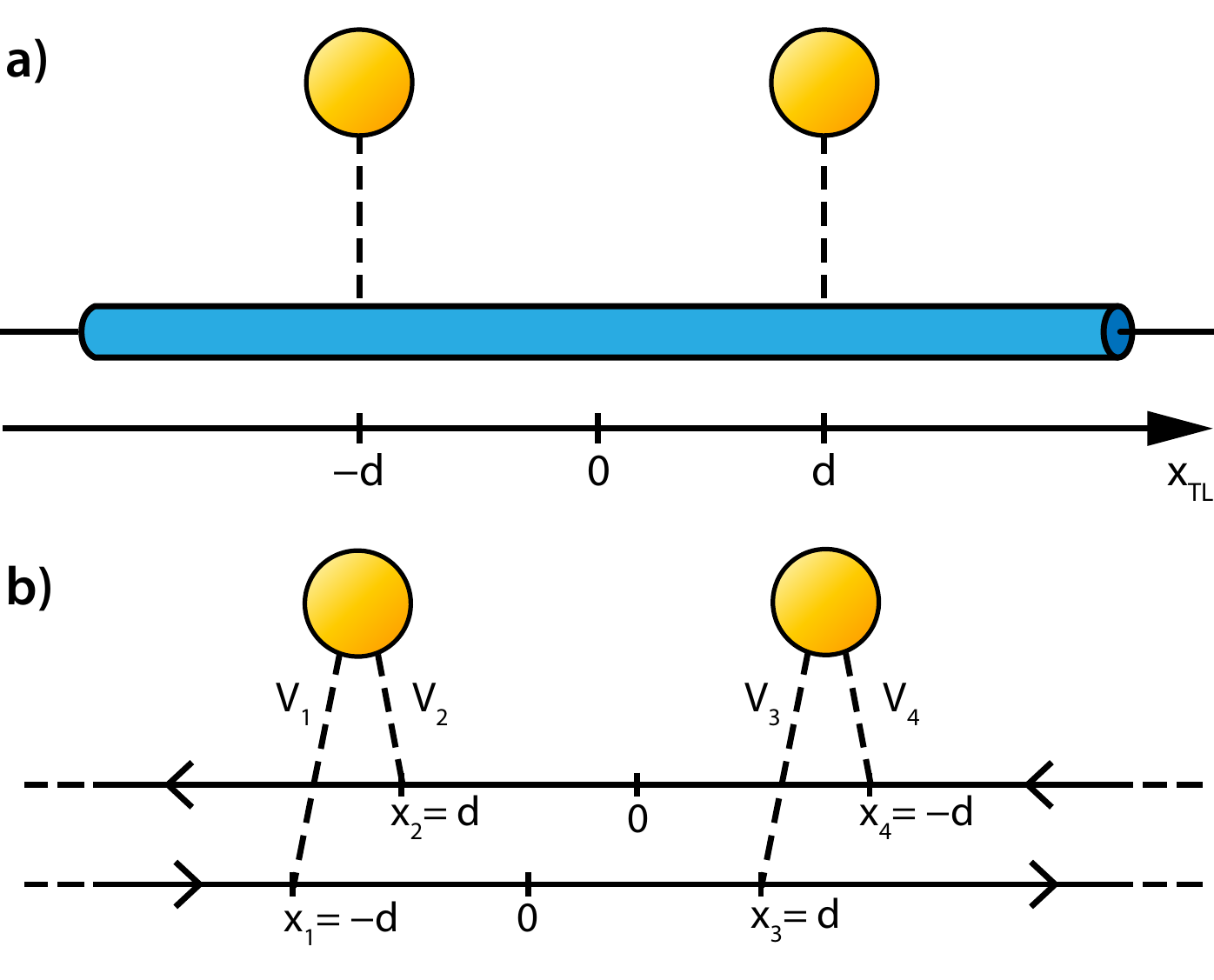}
  \caption{(Color online) Representation of left- and right-propagating modes of the bi-directional waveguide as a pair of two chiral channels. Each channel requires its own coordinate axis co-directed with the propagation of the photons.}
  \label{fig:chiralrepr}
\end{figure}

Existing photonic experiments often rely on one-dimensional waveguides which naturally accommodate two counter-propagating (left- and right-moving) one-dimensional channels. They can be modeled as a pair of chiral channels, as shown in Fig.~\ref{fig:chiralrepr}. This modeling requires the introduction of a separate coordinate axis for each channel, which is co-directed with the propagation direction of the photons. However, this does not reduce the complexity of the problem, since the representation in Fig.~\ref{fig:chiralrepr}(b) explicitly shows that the ordering of the coupling points along one channel is opposite to their ordering along the other channel.

Single chiral channels can also be realized as e.g. edge states in topological photonic lattice systems exhibiting the photonic spin-Hall effect~\cite{Plotnik2014}, and in systems of linear optical resonator arrays with an induced artificial magnetic field~\cite{Hafezi2013a}. Additionally, chiral propagation may emerge in cascaded systems like tapered fibers coupled asymmetrically to the two directions of propagation due the ``spin orbit effect for light''~\cite{Petersen2014}, or in photonic crystal lattices using similar effects~\cite{Sollner2015a}. Several theoretical proposals now rely explicitly on chiral quantum optics models~\cite{Hafezi2011a,Lodahl2017,Pichler2015}. Thus, by modeling the guiding of photons using chiral channels, we may describe both traditional waveguides as well as more exotic chiral propagation modes. 

We model a general chiral channel, $\sigma$, in the wide band limit where its 
dispersion $\omega_{\sigma k}$ has been linearized around the working frequency, $\omega_0$, such that ${\omega_{\sigma k} = v_{\sigma} (k-k_0) + \omega_0}$, $v_\sigma$ being the group velocity in the channel $\sigma$. Introducing the bosonic creation operator, $\create{a}{\sigma \omega} = (2\pi v_{\sigma})^{-1/2} \int_{-\infty}^\infty \id{x} e^{i (\omega_0 + \omega) x/v_{\sigma}} \create{a}{\sigma} (x)$, which creates a single photonic excitation with a frequency $\omega_0 + \omega$ in channel $\sigma$, we end up with the single channel Hamiltonian,
\begin{align}
  H_{\sigma} &= \int \id{\omega} (\omega_0 + \omega)\create{a}{\sigma \omega} \annihilate{a}{\sigma \omega}.
\end{align}
In the following we assume for simplicity that all $v_{\sigma}$ are the same in all channels, and use convenient units where ${\hbar= v_{\sigma} =1}$. In this way, $H_{\text{chs}} = \sum_\sigma H_\sigma$.

The dynamics of the chiral channels, $H_{\text{chs}}$, or the scatterer, $H_{\text{sc}}$, are solvable in themselves, and all of our difficulty arises from their mutual coupling, $V$. A single coupling term, $V_n$, connects a point, $x_n$, along a chiral channel, $\sigma_n$ to the $j_n$'th site on the scatterer with the overall coupling strength, $g_n$. We define it through the dipole coupling operator in the rotating wave approximation (RWA),
\begin{align}
  V_n &= g_n \create{b}{j_n} \annihilate{a}{\sigma_n}(x_n) + h.c.\nonumber \\
  &= \int \id{\omega} (g_n e^{i (\omega_0 + \omega) x_n} \create{b}{j_n} \annihilate{a}{\sigma_n, \omega} + h.c.).
  \label{eqn:Vn}
\end{align}
For the sake of brevity we also define the total effective coupling between one channel, $\sigma$, and the scatterer, 
\begin{align}
  V_\sigma &= \sum_n \delta_{\sigma, \sigma_n} V_n
           = \int \id{\omega} (\create{\tilde{b}}{\sigma} \annihilate{a}{\sigma, \omega} + h.c.).
\end{align}
In the last line we introduced the shorthand ${\create{\tilde{b}}{\sigma} = \sum_n \delta_{\sigma, \sigma_n} g_n e^{i (\omega_0 + \omega) x_n} \create{b}{j_n}}$ that describe the collective excitation originating from channel $\sigma$.

% section model (end)
\section{Diagrammatic scattering theory} % (fold)
\label{sec:diagrammatic_scattering_theory}

In this section we briefly remind the reader of diagrammatic Lippmann-Schwinger scattering theory and its application to photonic systems as derived in Refs.~\cite{Pletyukhov2012a, Pletyukhov2015}. We then apply diagrammatic scattering theory to our general photonic system; a Bose-Hubbard graph coupled to multiple chiral channels. 

% Again, our scattering setup consists of a number of open chiral one-dimensional channels, $H_{\text{chs}}$, that couple to a scatterer, $H_{\text{sc}}$, through linear couplings, $V$. 
% \begin{align}
%   H &= H_0 + V = H_{\text{sc}} + H_{\text{chs}} + V.
% \end{align}
Photons begin their journey in the channels far away from the scatterer. They propagate freely along their original channel, then encounter the scatterer and interact with it until finally leaving the scatterer through the (same or possibly other) channels.

This time-evolution from the initial photonic state at time $t_i$, to the final scattered state at time $t_f$ is defined by the time-evolution operator in the interaction picture, $U (t_f, t_i)= \mathcal{T} \exp(-i \int_{t_i}^{t_f} \id{\tau} V (\tau))$. Scattering theory considers the full time evolution from the distant past, $t_i\rightarrow-\infty$ where the incoming state is far away from the scatterer, and to the far future, $t_f\rightarrow+\infty$ where the outgoing state has not only left the scatterer but is also far away from it.

In this limit, the time-evolution operator is simply known as the $S$-matrix. When working explicitly in the interaction picture, it expresses scattering transitions between eigenstates of the uncoupled system, ${H_0 = H_{\text{chs}} + H_{\text{sc}}}$, under the physical constraint that the energy of the scattered state is conserved, $E_i = E_f$. 

The trivial part of the $S$-matrix can be separated out through the introduction of the $T$-matrix~\cite{Taylor2006},
\begin{align}
  S = I - 2 \pi i T (E_i) \delta(E_f - E_i).
\end{align}
The $T$-matrix can be perturbatively expanded in the interaction $V$,
\begin{align} \label{eqn:Tmatrix}
  T (E)
    &= V+ V \frac{1}{E - H_{0} - V + i 0^+} V \nonumber \\
    &= \sum_{j = 0}^\infty V \left(\frac{1}{E - H_{0} + i 0^+} V \right)^j \nonumber \\
    &= \sum_{j = 0}^\infty V \left( G_0 (E) V \right)^j,
\end{align}
and this expression forms the basis for the diagrammatic approach. In the last line we introduced the bare Green's function $G_0(E) = (E - H_0 + i 0^+)^{-1}$ that propagate the system in between consecutive scattering events. 

\begin{figure}[!ht]
  \centering
  \includegraphics[width=\columnwidth]{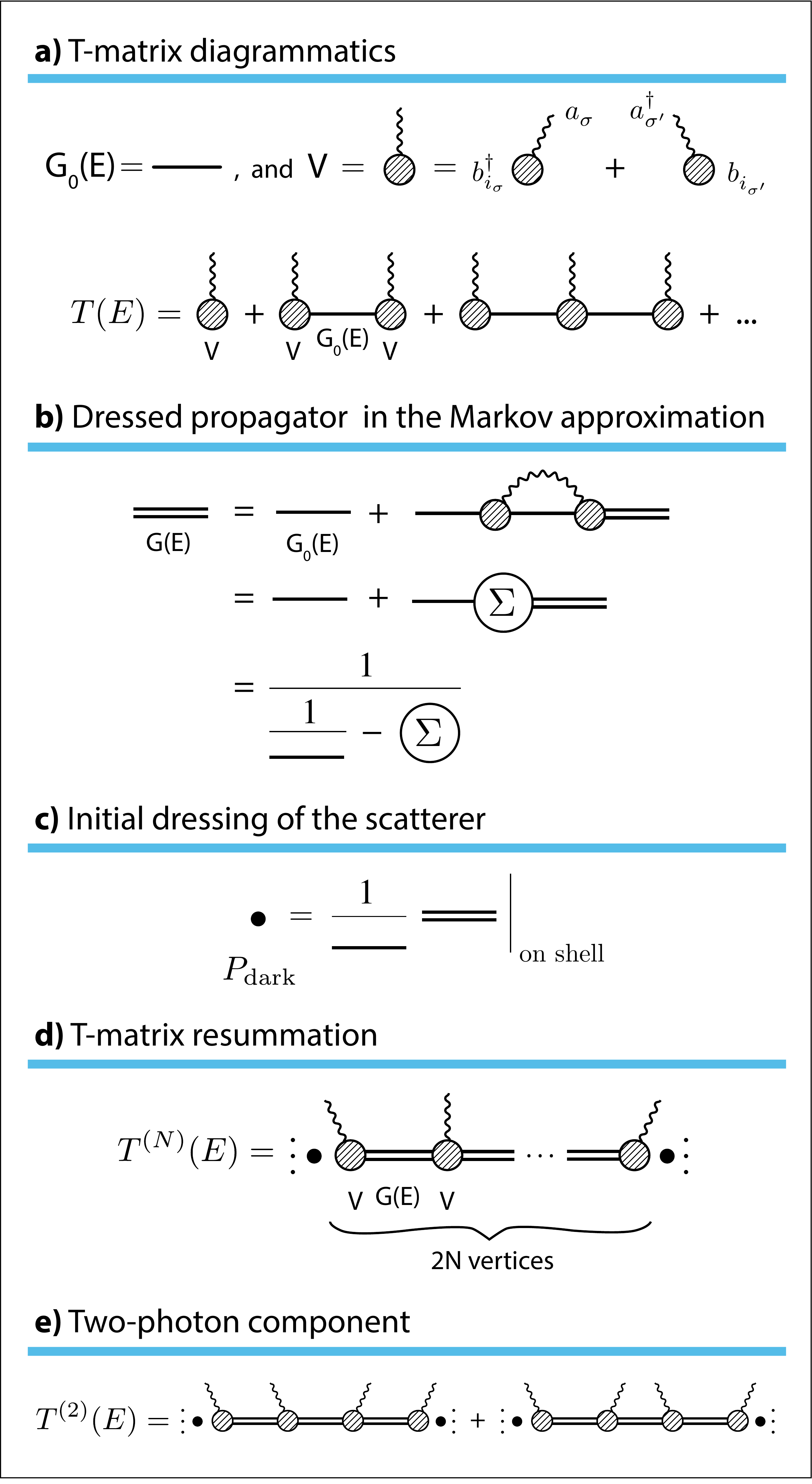}
  \caption{(Color online) Diagrammatic form of the scattering theory developed in Ref.~\cite{Pletyukhov2012a}. \textbf{(a)} The expression for $T$-matrix as derived in Eq.~\eqref{eqn:Tmatrix}. \textbf{(b)} Diagrammatic expression of the dressing of the bare Greens function by the emission and immediate re-absorption of virtual photons. \textbf{(c)} The dressing of the initial state of the scatterer, which effectively projects onto the dark state subspace. \textbf{(d)} The $N$-photon operator contribution to the $T$-operator. Notice the modified normal ordering, \footnotesize$\vdots \qquad \vdots$\normalsize, which signifies an additional shift to the Green's function arguments as described in the main text. \textbf{(e)} Diagrams for the two-photon $T$-operator.}
  \label{fig:diagrams}
\end{figure}

We draw a scattering event, $V$, as a filled circle with an additional wiggly lines pointing to the right (left) and symbolizing the absorption (emission) of a photon by the scatterer. The bare propagation, $G_0(E)$, is drawn as a solid line. The $T$-matrix in Eq.~\eqref{eqn:Tmatrix} can then be drawn as shown in Fig.~\ref{fig:diagrams}(a).

A photon that is emitted during a scattering event, $V$, can either belong directly to the outcoming state; or it can be a virtual photon which is later re-arbsorbed by the scatterer. Diagrammatic scattering theory deals with the problem of how to correctly describe these virtual photon processes.

\citet{Pletyukhov2012a} showed that under the assumption of two Markovianity conditions only a single type of virtual processes become important to scattering. The Markovianity conditions demand that
\begin{enumerate}
    \item all channels have linear dispersions, and
    \item dipole couplings are independent of the photonic mode, $\omega$;
\end{enumerate}
and they ensure that the whole scattering problem can be captured exactly within the Markov approximation. 

The only virtual processes that contribute to the scattering amplitudes under these conditions involve virtual photons which are re-absorbed immediately after their emission~\cite{Pletyukhov2012a}. These virtual processes are then be reincorporated as a dressing of the bare propagator, giving rise to a Dyson equation for the dressed Green's function as shown in Fig.~\ref{fig:diagrams}(b). 

Direct inversion of the the Dyson equation gives access to the dressed Green's function, $G(E)^{-1} = G_0(E)^{-1} - \Sigma (E)$, where $\Sigma (E)$ is the scatterer's self-energy which captures all possible consecutive processes involving the emission of a virtual photon and its immediate absorption. It equals
\begin{align} \label{eqn:Sigma}
\Sigma (E) &= \tilde{\Sigma} (E -H_{\text{chs}}), 
\end{align}
where
\begin{align}
	\tilde{\Sigma} (E) &= \left\langle V G_{0} (E) V \right\rangle_{\text{chs}},
	\label{eqn:tilSigma}
\end{align}
with the average performed in the vacuum state of the channels. 

Additionally, these virtual processes also lead to a projection of the initial and final state of the scatterer onto its dark state subspace, when the expression for the {$S$-matrix} is put on-shell (i.e. it respects energy conservation). Formally, this projection is given by the operator $P_{\text{dark}} = G_0^{-1} G |_{\text{on shell}} = G G_0^{-1}|_{\text{on shell}}$, which is depicted in Fig.~\ref{fig:diagrams}(c). Assuming that the empty scatterer is the \emph{only} state in the dark-state subspace, this projection -- along with the RWA -- ensures the conservation of the number of photons in the incoming and outgoing states. The $T$-matrix can then be split up into the sum $T = \sum_{N=1}^\infty T^{(N)}$, where the contribution
\begin{align} \label{eqn:TN}
  T^{(N)}(E) = P_{\text{dark}} \left\{ \vvdots \left( \prod_{i=1}^{2N-1} V G(E) \! \right) V \, \vvdots \right\} P_{\text{dark}}
\end{align}
represents the $N$-photon operator multiplied by the dark-state projector. Here, we introduced a modified normal ordering of the chiral channel field operators, $\vvdots (\cdots) \vvdots$, which differs from the conventional normal ordering by the implementation of additional shifts in the Green's functions energy argument, as dictated by ${G (E+\omega_0 + \omega) \create{a}{\sigma \omega} = \create{a}{\sigma \omega} G (E)}$ and ${\annihilate{a}{\sigma \omega} G (E) = G (E-\omega_0 - \omega) \annihilate{a}{\sigma \omega} }$. While obscure, this modified normal ordering presents a simplified way to deal with all the possible diagrams. A graphical representation of Eq.~\eqref{eqn:TN} is shown in Fig.~\ref{fig:diagrams}(d).

Note that this formulation of few photon scattering is almost equivalent the cluster form in input-output theory~\cite{Xu2015}.

% section diagrammatic_scattering_theory (end)
\subsection{Local scatterers} % (fold)
\label{sub:local_scatterers}
The main obstacle in directly applying diagrammatic scattering theory to the relevant models, is that the coupling in Eq.~\eqref{eqn:Vn}, depends explicitly on the mode, $\omega$, and hence does not fulfill the second Markovianity condition. One way around this, is to limit ourselves to \emph{local} scatterers, where any given chiral channel, $\sigma$, only couples to the scatterer at a single point $x_\sigma$ along the channel. The coordinate of the coupling point can then be \emph{gauged} away, i.e. by setting $x_\sigma = 0$ in the expression for the coupling elements in Eq.~\eqref{eqn:Vn}. This allows us to write the couplings as
% in Eq.~\eqref{eqn:Vn} as
\begin{align}
  V_n &= \int \id{\omega} (g_n \create{b}{j_n} \annihilate{a}{\sigma_n, \omega} + h.c.).
\end{align}
Both Markovianity conditions can then be satisfied, and the diagrammatic approach of \citet{Pletyukhov2012a} allows us to solve the scattering problem exactly for an arbitrary number of incoming of photons. This subsection shows the derivation of the single photon and two-photon $S$-matrices for locally coupled scatterers.

%The total Markov treatment is given in~\citet{Lehmberg1970a,Lehmberg1970}

The first step in this derivation concerns the scatterer self-energy as defined by Eq.~\eqref{eqn:Sigma}. The self-energy is naturally split up in its contributions from any two coupling terms, $\Sigma = \sum_{n,m} \Sigma_{nm}$, where
\begin{align} \label{egn:Sigmasigma}
  \Sigma_{nm} &= \tilde{\Sigma}_{nm} =
    \left\langle V_n \frac{1}{E - H_0 + i 0^+} V_m \right\rangle_{\text{chs}}
  \nonumber \\
    &=
      g_n g_m^* \create{b}{j_{n}} \int_{-\infty}^\infty \id \omega \frac{1}{E - \omega_0 - \omega - H_{\text{sc}} + i 0^+} \annihilate{b}{j_m}
  \nonumber \\ 
    &= - i \pi g_n g_m^* \delta_{\sigma_n,\sigma_m} \create{b}{j_{n}} \annihilate{b}{j_m}.
\end{align}

The total effective scatterer Hamiltonian is the concatenation of the bare scatterer Hamiltonian and the self-energy, $H_{\text{eff}} = H_{\text{sc}}+ \Sigma$. Both the scatterer Hamiltonian and the self-energy conserve the number of photons in the scatterer, $M = \sum_{j} b^\dagger_{j} b_j$, i.e. $[ H_{\text{sc}}, M]  = [\Sigma, M] = 0$. Hence we can perform a spectral decomposition of the effective scatterer Hamiltonian for every photon number $m$, which is the eigenvalue of $M$. Since the self-energy is non-Hermitian, this decomposition,
\begin{align}
  H^{(M)}_{\text{eff}} &= \sum_{l_M=1}^{d_M} \lambda_{l_M}^{(M)} | M, l_M \rangle \langle \overline{m, l_M} | \nonumber\\ &= \sum_{l_M=1}^{d_M} \lambda_{l_M}^{(M)} P^{(M)}_{l_M},
  \label{eqn:eigendecomp}
\end{align}
contains both right $|m, l_m\rangle$ and left $\langle \overline{m, l_m}|$ eigenvectors of $H_{\text{eff}}^{(M)}$, corresponding to the (possibly complex) eigenvalue $\lambda_{l_M}^{(M)}$ and obeying the bi-orthogonality relations $\langle \overline{M, l_M}|M, l'_M\rangle = \delta_{l_M, l'_M}$. Here, $d_m$ denotes the dimension of the $M$-photon scatterer subspace, and $P^{(M)}_{l_M}$ is a projector onto the eigenspace of the eigenvalue $\lambda_{l_M}^{(M)}$. 

The corresponding dressed Green's function has the form
\begin{align}
G^{(M)} (E) &= \frac{1}{E - H_{\text{chs}} - H^{(M)}_{\text{eff}}} \nonumber \\ &= \sum_{l_M =1}^{d_M} \frac{P^{(M)}_{l_M}}{E - H_{\text{chs}} - \lambda_{l_M}^{(M)}}.
\label{GMlocal}
\end{align}
Note that it is important to keep $H_{\text{chs}} $ in this definition in order to maintain the diagrammatic rules discussed in the paragraph following Eq.~\eqref{eqn:TN}.

The single photon $S$-matrix describes how an incoming single photon state, $| \nu_1 \rangle = \create{a}{\nu_1}\ket{0}_{\text{chs}}$, scatters into multiple outgoing single photon states, $|\nu'_1\rangle$. Note that we refer to photonic states in the channels using a multi-index, $\nu = (\sigma, \omega)$, labelling both the channel, $\sigma$ and the mode, $\omega$. We may write ${S^{(1)} = \sum_{\nu'_1, \nu_1} S^{(1)}_{\nu'_1, \nu_1} \create{a}{\nu'_1} \annihilate{a}{\nu_1}}$, where 
\begin{align}
  S^{(1)}_{\nu'_1, \nu_1} &= S^{(1)}_{\sigma'_1 \omega'_1, \sigma_1 \omega_1} \nonumber \\
  &= \delta_{\omega'_1 \omega_1} \left( \delta_{\sigma'_1 \sigma_1} - 2 \pi i T^{(1)}_{\sigma'_1 \omega'_1, \sigma_1 \omega_1} \right).
  \label{S1me}
\end{align}
Applying Eq.~\eqref{eqn:TN}, we find
\begin{align*}
  T^{(1)} &= P^{(0)} \, \vvdots \, V G V \, \vvdots \, P^{(0)}
    \\ &= \sum_{\sigma'_1 \omega'_1, \sigma_1 \omega_1} 
    P^{(0)} \annihilate{\tilde{b}}{\sigma'_1} G^{(1)} (E + \omega_0 + \omega'_1) \create{\tilde{b}}{\sigma_1} P^{(0)} \nonumber \\
    & \qquad \qquad \qquad \times \create{a}{\sigma'_1 \omega'_1}  \annihilate{a}{\sigma_1 \omega_1}.
\end{align*}
Enforcing the on-shell condition, ${\omega'_1 = \omega_1}$, and applying the spectral decomposition to the single-photon dressed Green's function, we obtain an expression for the matrix elements appearing in Eq.~\eqref{S1me},
\begin{align} \label{eqn:T1}
  T^{(1)}_{\sigma'_1 \omega'_1, \sigma_1 \omega_1} 
    &=P^{(0)}  \sum_{l_1 =1}^{d_1} 
        \frac{\langle 0 | \annihilate{\tilde{b}}{\sigma'_1} | 1, l_1 \rangle \langle \overline{1, l_1} | \create{\tilde{b}}{\sigma_1} | 0 \rangle}
        {\omega_1 + \omega_0 - \lambda_{l_1}^{(1)}} .
\end{align}
 
% The resulting scattering amplitude squared is identical to the Landauer-B\"{u}ttiker transmission~\cite{Landauer1957, Markussen2010}known from quantum transport through electronic systems.

The effects of photonic correlations first show up in the scattering of two-photon states $|\nu_1, \nu_2\rangle = \create{a}{\nu_1} \create{a}{\nu_2} \ket{0}_{\text{chs}}$ into other two photon states, $|\nu'_1, \nu'_2\rangle$. One can generally show \cite{Pletyukhov2012a,Laakso2014} that the two-photon scattering matrix
\begin{align} \label{eqn:S2}
  S^{(2)}
  &= \sum_{\nu'_1, \nu'_2} \sum_{\nu_1, \nu_2}  \create{a}{\nu'_1} \create{a}{\nu'_2}
      \annihilate{a}{\nu_2} \annihilate{a}{\nu_1}  \\
  &\times \left(
    \frac{1}{2} S_{\nu'_1 , \nu_1}^{(1)} S_{\nu'_2 , \nu_2}^{(1)} 
    - 2 \pi i T^{(2) \mathcal{P}}_{\nu'_1, \nu'_2; \nu_1, \nu_2 } \delta_{\omega'_1 + \omega'_2, \omega_1 + \omega_2}
    \right)  \nonumber
\end{align}
can be expressed in terms of the single-photon matrices from Eq.~\eqref{S1me} and the principal value part ($\mathcal{P}$) of the two-photon $T$-matrix calculated on the basis of Eq.~\eqref{eqn:TN}. The first contribution describes elastic scattering  of two photons (i.e. without energy exchange between them), while the second contribution captures inelastic effects arising due to the effective interaction between the two photons. 

A direct calculation of the two-photon inelastic matrix components, which is based on the diagrams in Fig.~\ref{fig:diagrams}{(e)} and presented in Appendix~\ref{app:local}, yields the following result
\begin{widetext}
\begin{align}
  T^{(2) \mathcal{P}}_{\nu'_1, \nu'_2; \nu_1, \nu_2 } &= 
    P^{(0)}  \sum_{l_1, l'_1}
    \frac{\langle 0| \annihilate{\tilde{b}}{\sigma'_1} | 1, l'_1 \rangle}{\omega'_1 + \omega_0 - \lambda_{l'_1}^{(1)}}  
    \Bigg\{ \sum_{l_2} \frac{ \langle \overline{1, l'_1} | \annihilate{\tilde{b}}{\sigma'_2} | 2 , l_2 \rangle \langle \overline{2, l_2} | \create{\tilde{b}}{\sigma_2} | 1, l_1 \rangle  }{\omega_1+ \omega_2 +2 \omega_0 - \lambda_{l_2}^{(2)}} 
    \nonumber \\ 
    & \qquad \qquad \qquad \qquad 
    - \frac{1}{2}\frac{\omega_1+\omega_2 +2 \omega_0 - \lambda_{l'_1}^{(1)} -\lambda_{l_1}^{(1)}}{ (\omega'_2 + \omega_0 - \lambda_{l_1}^{(1)}) (\omega_2 + \omega_0 - \lambda_{l'_1}^{(1)}) } \langle \overline{1, l'_1} | \create{\tilde{b}}{\sigma_2}  | 0 \rangle \langle 0 |\annihilate{\tilde{b}}{\sigma'_2} | 1, l_1 \rangle  \Bigg\} 
    \frac{\langle \overline{1, l_1} | \create{\tilde{b}}{\sigma_1} | 0 \rangle}{\omega_1+ \omega_0 - \lambda_{l_1}^{(1)}}.
    \label{eqn:T2}
\end{align}
\end{widetext}
This expression provides an exact description of two-photon scattering, and while it is unwieldy for all but the simplest scatterers, it does allow for an efficient numerical calculation of photonic correlation effects showing up in two photon scattering. 
% subsection local_scatterers (end)
\subsection{Quasi-local scatterers} % (fold)
\label{sub:quasi_local_scatterers}
In the previous section we ensured the fulfillment of the Markovianity conditions by restricting ourselves to local scatterers, and this approach provided us with both a simple and exact solution of the scattering problem. 

We now turn to systems where the channels couple non-locally to the scatterer, such that coupling points on some channel are separated by distances greater than the relevant wave-length, $\omega_0 |x_n - x_m| \gtrsim 1$. The dynamics of such systems is in general non-Markovian.

However, there exists a quasi-local regime, where we can still obtain trustworthy results by applying Markov-type simplifications to the dynamics. In this regime, the non-Markovian effects originating from higher order scattering diagrams, which renormalize bare vertices and therefore lie beyond the set of diagrams shown in Fig.~\ref{fig:diagrams}, are perturbatively small, as long as the condition 
\begin{align}
D_{nm} = \pi |g_n g_m| |x_{n} - x_{m}| \ll 1
\label{mark_cond1}
\end{align}
 is fulfilled. This condition means that the scatterer's state is unlikely to decay on time scales of a field propagation between two coupling points. In addition to Eq.~\eqref{mark_cond1} we also assume that
\begin{align}
|| M \omega_0  - H_{\text{sc}}^{(M)}|| \cdot |x_n - x_m| \ll 1, \text{ for all }m
\label{mark_cond2}
\end{align}
This condition implies that the scatterer's internal energy scales -- as characterized by its detunings and non-linearities -- do not substantially contribute to the propagation phase on a typical distance between two coupling points. In other words, the scatterer dynamics in the co-rotating frame is frozen during the field propagation from one coupling point to the next one.

The insignificance of non-Markovian effects in the regime of parameters restricted by the conditions in Eq.~\eqref{mark_cond1} and Eq.~\eqref{mark_cond2} has been explicitly checked in Ref.~\cite{Laakso2014} for a model of two two-level systems coupled to a bi-directional waveguide at different points by considering the quasi-local limit of the exact analytic non-Markovian solution derived for this model. 

The Markovian approximation in the quasi-local case -- relying on the conditions Eq.~\eqref{mark_cond1} and Eq.~\eqref{mark_cond2} -- is achieved on the basis of the same diagrams, shown in Fig.~\ref{fig:diagrams}, as in the local case. However, it is necessary to additionally specify the values of their constituent dressed Green's functions and bare vertices.

At the outset, a model with a quasi-local coupling--in contrast to its local counterpart--allows for the coupling to have mode-dependent phase factors, as appearing in Eq.~\eqref{eqn:Vn}. This dependence is essential in calculating the self-energy from Eq.~\eqref{eqn:tilSigma}. It leads to the dependence of $\tilde{\Sigma}_{nm}$ on the ordering of the coupling points along the coordinate axis of the relevant channel. In particular,
\begin{align}
 \tilde{ \Sigma}_{nm} (E) &= 
    \left\langle V_n \frac{1}{E - H_0 + i 0^+} V_m \right\rangle_{\text{chs}}
  \nonumber \\
  &=\delta_{\sigma_n, \sigma_m} g_n g_m^* \nonumber \\
  & \times \create{b}{j_n} \left( \int_{-\infty}^\infty \id{\omega}  \frac{e^{i (\omega_0 + \omega ) (x_n - x_m)} }{E - \omega-\omega_0 - H_{\text{sc}} + i 0^+} \right)\annihilate{b}{j_m}
  \nonumber \\
  &= - 2 i \pi g_n g_m^*  \delta_{\sigma_n, \sigma_m}\Theta(x_n - x_m) \nonumber \\
  &\qquad \times
    \create{b}{j_n} e^{i ( E - H_{\text{sc}})(x_n - x_m)} \annihilate{b}{j_m}  .
    \label{eqn:tilSigmanM}
\end{align}
Here the Heaviside step function is conveniently defined such that $\Theta(0) = \frac{1}{2}$. In the derivation of Eq.~\eqref{eqn:tilSigmanM} we made use of a linear channel dispersion. Mathematically the difference between the two possible orderings of the coupling points stems from the condition whether the phase factor allows us to close the integration contour in either the upper or lower complex half-plane.

According to Eq.~\eqref{eqn:Sigma} we find the dressed Green's function of the scatterer
\begin{align}
G (E) =  \frac{1}{E - H_{\text{chs}} - H_{\text{sc}} - \tilde{\Sigma} (E- H_{\text{chs}})},
\label{Gqlocal}
\end{align}
where $\tilde{\Sigma} = \sum_{n,m} \tilde{\Sigma}_{nm}$. It can be also decomposed into a sum of contributions with different photon numbers $M$,
\begin{align}
G^{(M)} (E) =  \frac{1}{E - H_{\text{chs}} - H_{\text{sc}}^{(M)} - \tilde{\Sigma}^{(M)} (E- H_{\text{chs}})}.
\label{GMqlocal}
\end{align}
We see that Eq.~\eqref{GMqlocal}, however, does not have the Markovian form inherent to its local counterpart Eq.~\eqref{GMlocal}: because of the functional dependence of $\tilde{\Sigma}^{(M)}$ on $E- H_{\text{chs}}$ the Green's function in Eq.~\eqref{GMqlocal} might have more than $d_M$ poles.

We can enforce the Markov approximation for quasi-local scatterers, assuming that the initial state consists of $N$ photons, all having the same frequency $\omega_0$. Then, the approximation
\begin{align}
G^{(M)} (N \omega_0 )  \approx  \frac{1}{N \omega_0 - H_{\text{chs}} - H_{\text{sc}}^{(M)}  - \tilde{\Sigma}^{(M)} (M \omega_0)} 
\label{GMappr}
\end{align}
allows us to introduce the effective scatterer Hamiltonian $H_{\text{eff}}^{(M)} = H_{\text{sc}}^{(M)}  + \tilde{\Sigma}^{(M)} (M \omega_0)$ and to finally bring Eq.~\eqref{GMqlocal} to the form of Eq.~\eqref{GMlocal}. 

We note that the approximation in Eq.~\eqref{GMappr} is consistent with the quasi-local condition of Eq.~\eqref{mark_cond1}. Indeed, expanding $\tilde{\Sigma}^{(M)} (E- H_{\text{chs}})$ around $M \omega_0$ up to a linear term, we obtain a correction to $E-  H_{\text{chs}} - H_{\text{sc}}^{(M)}$ in Eq.~\eqref{GMlocal}, which is smaller by a factor of the order of $D_{nm}$ than the leading term.

Using the condition in Eq.~\eqref{mark_cond2}, we can additionally simplify $\tilde{\Sigma}^{(M)} (M \omega_0)$. From Eq.~\eqref{eqn:tilSigmanM} it follows
\begin{align}
\tilde{\Sigma}^{(M)}_{nm} (M \omega_0) &= - 2 i \pi g_n g_m^*  \delta_{\sigma_n, \sigma_m}\Theta(x_n - x_m) \nonumber \\
  &\qquad \times
    \create{b}{j_n} e^{i ( M \omega_0 - H^{(M-1)} _{\text{sc}})(x_n - x_m)} \annihilate{b}{j_m}  \nonumber \\
    & \approx  - 2 i \pi g_n g_m^*  \delta_{\sigma_n, \sigma_m}\Theta(x_n - x_m) \nonumber \\
  &\qquad \times
    \create{b}{j_n}  \annihilate{b}{j_m}  e^{i \omega_0 (x_n  - x_m)}.
    \label{eqn:tilSigmanMappr}
\end{align}

Thus, computing a scattering matrix for a quasi-local scatterer is possible on the basis of the diagrams in Fig.~\ref{fig:diagrams} and the associated diagrammatic rules described in the previous subsection. Additionally, the dressed Green's functions must be approximated by Eq.~\eqref{GMappr}, where the self-energies are given by Eq.~\eqref{eqn:tilSigmanMappr} instead of Eq.~\eqref{egn:Sigmasigma}. Finally, bare vertices in the diagrams of Fig.~\ref{fig:diagrams}{(d,e)} must be replaced by
\begin{align} \label{eqn:CouplingQLproj}
  V_n  \to \tilde{V}_n = \int \id{\omega} (g_n e^{i \omega_0 x_n} \create{b}{j_n} \annihilate{a}{\sigma_n , \omega} + h.c.).
 \end{align}
The reason for neglecting the phase factor $e^{i \omega x_n}$ in the above expression is based on the observation that the diagrams in Fig.~\ref{fig:diagrams} produce poles in the functional dependence of $T$ matrix on frequencies $\omega'_1, \omega'_2, \ldots$, which are located at points $\lambda_{l_M}^{(M)} -M \omega_0$ in the lower half of the frequency complex plane (see e.g. Eq.~\eqref{eqn:T2}). Therefore, the contribution of the aforementioned phase factors can be estimated by $\exp(- i (\lambda_{l_M}^{(M)} -M \omega_0) |\Delta x|) \approx 1$, in accordance with Eq.~\eqref{mark_cond1} and Eq.~\eqref{mark_cond2}. Note that these two quasi-locality conditions can be combined into a single one
\begin{align}
|\lambda_{l_M}^{(M)} -M \omega_0|  |\Delta x| \ll 1.
\end{align}
% subsection quasi_local_scatterers (end)
\subsection{Observables} % (fold)
\label{sub:observables}
% Scattering theory describes scattering of an arbitrary incoming photonic state.
Since scattering matrices are naturally expressed in terms of the non-normalizable eigenbases of the channels, special care must be taken in applying scattering theory to real, physical states. We construct such physical incoming states as rectangular wave packets quantified by their width $L$, central coordinate $x_c$, channel $\sigma_0$, and working frequency $\omega_0$
\begin{align}
  \create{\mathcal{A}}{\sigma_0, \omega_0; x_c} &= \frac{1}{\sqrt{L}}  \int_{x_c-L/2}^{x_c + L/2} \id{x} \create{a}{\sigma_0} (x) e^{i \omega_0 x} \nonumber \\
  &= \int_{-\infty}^{\infty} {\id{\omega}} e^{-i \omega x_c} \phi(\omega ) \create{a}{\sigma_0 \omega}.
\end{align}
In the interaction picture, in which the scattering formalism is formulated, we have
\begin{align}
e^{i H_0 t}  \create{\mathcal{A}}{\sigma_0, \omega_0; x_c} e^{-i H_0 t} = e^{i \omega_0 t} \create{\mathcal{A}}{\sigma_0, \omega_0; x_c -t} .
\label{wpint}
\end{align}
The convolution function, ${\phi(\omega) = \sqrt{L/(2\pi)}\, \mathrm{sinc}(\omega L /2)}$, approaches $ \sqrt{2 \pi/L} \delta (\omega)$ in the wide packet limit, ${L\rightarrow \infty}$. Therefore Eq.~\eqref{wpint} can be approximated by $e^{i \omega_0 t} \create{\mathcal{A}}{\nu_0} = \create{\mathcal{A}}{\nu_0} (t) $, where $\create{\mathcal{A}}{\nu_0}$ is the normalized wavepacket operator,
\begin{align}
\create{\mathcal{A}}{\nu_0} = \int_{-\infty}^{\infty} {\id{\omega}}  \phi(\omega ) \create{a}{\sigma_0 \omega} , \quad \int_{-\infty}^{\infty} {\id{\omega}}  |\phi(\omega ) |^2 =1.
\end{align}

We focus in the following exclusively on the scattering of weakly coherent states.
Creating an incoming coherent state, which is written in the interaction picture as
\begin{align}
  | \phi_{\nu_0} \rangle &= e^{-\bar{n}/2} e^{-\sqrt{\bar{n}} \create{\mathcal{A}}{\nu_0} (t)} |0 ,0 \rangle ,\quad  |0,0\rangle =   |0 \rangle_{\text{chs}} |0  \rangle,
\end{align}
we assume that the mean number of photons $\bar{n}$ is small, and perform an expansion of $ | \phi_{\nu_0} \rangle$ in this parameter. The resulting scattering state $ | S \phi_{\nu_0} \rangle$ is given by the scattering matrix, and up to two-photon contribution we find
\begin{align}
  & | S \phi_{\nu_0} \rangle \nonumber \\
  &\approx e^{-\bar{n}/2} \left(
    1 + \sqrt{\bar{n}} S^{(1)} \create{\mathcal{A}}{\nu_0} (t) 
    +\frac{\bar{n}}{2} S^{(2)} (\create{\mathcal{A}}{\nu_0} (t))^2  
  \right) \ket{0,0}.
\end{align}
The properties of this state can be characterized through the measurement of the first and second order correlation functions. 

The first order correlation function is given by
\begin{align} \label{eqn:g1_1}
 g^{(1)}_{\sigma_0, \sigma} (\tau) &
    = \langle S \phi_{\nu_0} | \create{a}{\sigma} (x-t-\tau) \annihilate{a}{\sigma} (x-t) | S \phi_{\nu_0}\rangle ,
\end{align}
where the operators $\create{a}{\sigma} $, $\annihilate{a}{\sigma}$ are also written in the interaction picture. To leading order in $\bar{n}$, it can be expressed
\begin{align}
  g^{(1)}_{\sigma} (\tau)
  &\approx 
  \bar{n} \langle 0,0| \annihilate{\mathcal{A}}{\nu_0} (t) S^{(1) \, \dagger} \create{a}{\sigma} (x-t- \tau) \nonumber \\
   & \qquad \qquad \quad \times \annihilate{a}{\sigma} (x -t) S^{(1)} \create{\mathcal{A}}{\nu_0} (t) | 0,0\rangle 
  \nonumber \\
  &\approx
  f e^{i \omega_0 \tau} | s^{(1)}_{\sigma \sigma_0}|^2,
\end{align}
where $f = \bar{n}/L$ is the photonic flux and 
\begin{align}
s_{\sigma_0, \sigma}^{(1)} = \delta_{\sigma \sigma_0} - 2 \pi i \sum_{l_1 =1}^{d_1} 
        \frac{\langle 0 | \annihilate{\tilde{b}}{\sigma} | 1, l_1 \rangle \langle \overline{1, l_1} | \create{\tilde{b}}{\sigma_0} | 0 \rangle}
        {\omega_0 - \lambda_{l_1}^{(1)}},
        \label{s1sigma}
\end{align}
originates in Eq.~\eqref{S1me} and Eq.~\eqref{eqn:T1}. The components of Eq.~\eqref{s1sigma} contain the single-photon transmission ($\sigma \neq \sigma_0$) or reflection ($\sigma =\sigma_0$) amplitudes. This expression is general and well-known from any single particle transport theory, and due to the Fisher-Lee relation it is identical to the Landauer-B\"uttiker scattering matrix~\cite{Landauer1957}. Unless stated otherwise we discuss $g_{\sigma \sigma_0}^{(1)}(0)$, i.e. does not include the phase factor.

Interaction effects within two-photon scattering are available to experiments through the second order intensity correlation function, measurable in a Hanbury Brown and Twiss setup~\cite{HanburyBrown1956}, where the time offset, $\tau$, is directly translatable into real-space detection points, $\Delta x$, due to our assumption of a linear channel dispersion. The correlation of the scattered state is directly given by,
\begin{align} \label{eqn:g2}
  g^{(2)}_{\sigma_0 \sigma_0, \sigma \sigma'} (\tau)
    &= \frac{1}{g_{\sigma}^{(1)} (0) g_{\sigma'}^{(1)} (0)} \langle S \phi_{\nu_0} | \create{a}{\sigma'}(x-t) \create{a}{\sigma} (x -t- \tau) 
    \nonumber \\ 
    & \phantom{=} \cdot \annihilate{a}{\sigma} (x - t- \tau)\annihilate{a}{\sigma'} (x-t) | S \phi_{\nu_0} \rangle .
\end{align}
% To indicate the input channel, we also introduce an extended notation $g^{(2)}_{\sigma \sigma', \sigma_0 \sigma_0} (\tau)$ for this function.

Appendix~\ref{app:g2} details the analytical calculation of the $g^{(2)}$ correlation function for complex scatterers. The intensity-intensity correlation function probes the effects of photon-photon interaction present  within the scattering structure: Such correlations disappear in the absence of such photon-photon interactions ($g^{(2)}=1$), while the presence of any photon-photon interactions may either lead to photon bunching ($g^{(2)} > 1$) or photon anti-bunching ($g^{(2)} < 1$).

When the $g^{(1)}$ or $g^{(2)} $correlation functions involves different incoming and outgoing channels, their analytical expressions can be written very compactly. Specifically the expressions in Eq.~\eqref{eqn:g1_1} and Eq.~\eqref{eqn:g2_app_res} simplify to,
% Define effective $\tilde{H} = H_0 + \Sigma - \hat{N} \omega_0$
\begin{align}
  g_{1,2}^{(1)} (\tau) &= f e^{i \omega_0 \tau} \left|A_{1,2}^{(1)}(0)\right|^2, 
  \label{eqn:g1_2}
  \\
  g_{11,22}^{(2)} ( \tau )
    &=
    \left | 1 -  \frac{A_{1,2}^{(1)}(\tau)}{A_{1,2}^{(1)}(0)} + \frac{A_{11,22}^{(2)} (\tau)}{( A_{1,2}^{(1)}(0) )^2 } \right|^2, \label{eqn:g2_2}
\end{align}
Here the one- and two-photon amplitudes are given by,
\begin{align}
  A_{1,2}^{(1)}(\tau) &= \langle 0 | \annihilate{\tilde{b}}{2} e^{i (\omega_0 - H_{\text{eff}}) \tau} G^{(1)} \create{\tilde{b}}{1} |0 \rangle, \label{eqn:A1}
  \\
  A_{11,22}^{(2)}(\tau) &= \langle 0 | \annihilate{\tilde{b}}{2} e^{i (\omega_0 - H_{\text{eff}}) \tau} \annihilate{\tilde{b}}{2} G^{(2)} \create{\tilde{b}}{1} G^{(1)} \create{\tilde{b}}{1} | 0 \rangle. \label{eqn:A2}
\end{align}

Here, it is worth noting that $g^{(2)}$ is a relative measure of correlation, and strong correlation effects in $g^{(2)}$ are caused by dramatic changes happening in either the $A^{(1)}$ amplitude in the denominator of Eq.~\eqref{eqn:g2_2} or the two-photon amplitude $A^{(2)}$ in the numerator of Eq.~\eqref{eqn:g2_2}. In the results section we find that dramatic two photon correlation effects are most often associated with destructive interference effects in the single photon amplitude, $A^{(1)}$ leading to (relatively speaking) strong bunching of photons. Strong anti-bunching ($g^{(2)} < 1$) is the less common (hence, also more interesting) effect to achieve since it is most often associated with destructive two-photon interference phenomena (as expressed by the $A^{(2)}$ numerator of Eq.~\eqref{eqn:g2_2}).
% subsection observables (end)

\section{Results} % (fold)
\label{sec:results}
While the above expression can be evaluated analytically for simple systems, our final goal is to study complex extended scattering structures. To this end we have implemented the single and two photon correlation functions numerically in the Python programming language using the numpy package. The babusca code -- made available in the supplementary material -- is modular and flexible and makes it possible to construct arbitrary scattering systems composed of multiple chiral channels coupled to a Bose-Hubbard scatterer. For weakly coherent incident photons the code computes the single and two-photon correlation functions in the Markov approximation for systems with up to approximately a hundred sites on current standard equipment.

% We show and analyze the results of our calculation for a variety of simple uniform Bose-Hubbard lattices. However, since the lattice geometry and the lattice parameters may be chosen without restrictions, most of the (impossibly) large parameter space remains unexplored. 

As a pedagogical prelude, we consider transport through a single non-linear cavity; then turn to the dimer, which we discuss thoroughly, including an example of quasi-local coupling. We then investigate chain and ring geometries, and then finish by demonstrating how the scattering method also easily deals with larger scatterers by considering  an 8 by 8 square lattice penetrated by an artificial magnetic field and supporting quantum Hall edge states.

We fix the channel couplings of the incoming and outgoing channels, $\sum_\sigma \tfrac{1}{2} \Gamma_\sigma = \sum_\sigma \tfrac{1}{2} \pi g_\sigma^2 = 1$, effectively expressing the Bose-Hubbard parameters in units of this channel coupling. Because the Bose-Hubbard hopping amplitudes, $t$, determine the relevant bandwidth of the scatterer, we may in general associate strong channel coupling with weak hopping, $t \sim \Gamma$, and weak channel coupling with strong hopping, $t \sim 10 \Gamma$.

For each lattice we compute both the single-photon and two-photon correlations for weakly coherent incident photonic states, and plot the results as a function of the two-photon detuning, $2 \delta = 2 (\omega_0 - \langle \ve \rangle)$, where the averaged onsite energy $\langle \ve \rangle= 1/N_s \sum_{i=1}^{N_s} \ve_i$.

\subsection{The Kerr non-linear element} % (fold)
\label{sub:the_kerr_non_linearity}

We begin with a discussion of the simplest possible scattering setup: a single Bose-Hubbard site coupled symmetrically to two chiral channels. The scatterer Hamiltonian,
\begin{align}
H_{\text{sc}} = \varepsilon b_1^{\dagger} b_1 + \frac{U}{2} b_1^{\dagger} b_1 (b_1^{\dagger} b_1 - 1),
\end{align}
is characterized by a single resonance frequency, $\varepsilon$, and the photon-photon interaction, $U$. 

Due to its simplicity, this scattering setup have been realized on several experimental platforms by e.g. coupling a non-linear cavity to a one-dimensional transmission line~\cite{Astafiev2010}. Consequently, the analytical form of the coherence functions are well-known when driving the setup at both high and low power~\cite{Drummond1981,Shen2007,Liao2010,Pletyukhov2012a}. Thus, our goal in this section is pedagodical: to show how to derive the coherence functions in diagrammatic scattering theory, and to interpret the physics behind these results. 

Our single site scatterer is locally coupled, and its effective Hamiltonian, $H_{\text{eff}} = H_{\text{sc}} + \Sigma$, trivially supports a single (right) eigenstate in each $M$-photon sector, $|\lambda^{(M)}\rangle$, with eigenvalue, $\lambda^{(M)}$, as given by
\begin{align}
&|\lambda^{(0)}\rangle = \ket{0}, && \lambda^{(0)} = 0, \\
&|\lambda^{(1)}\rangle = \create{b}{1}\ket{0}, && \lambda^{(1)} = \varepsilon - i \Gamma, \\
&|\lambda^{(2)}\rangle = \tfrac{1}{\sqrt{2}}(\create{b}{1})^2\ket{0}, && \lambda^{(2)} = 2\varepsilon + U - 2 i \Gamma.
\end{align}

\begin{figure}[!htb]
  \centering
  \includegraphics[width=\columnwidth]{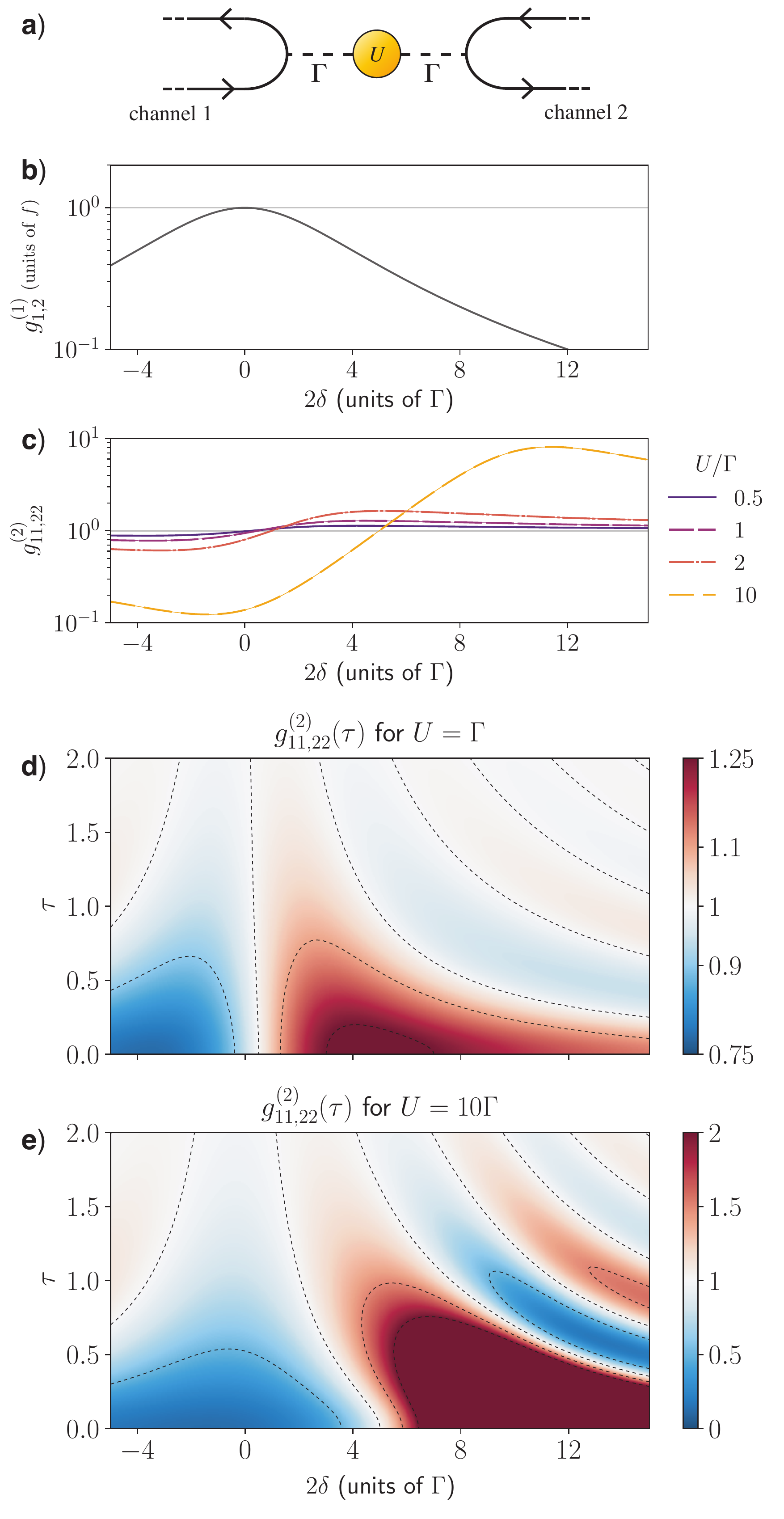}
  \caption{(Color online) A single Bose-Hubbard site with an interaction, $U$, coupled to two chiral channels. (\textbf{a}) The Lorentz shaped $g^{(1)}_{12}$ correlation (or transmission) between channel $1$ and $2$, with a resonant peak around zero detuning, $\delta = 0$. (\textbf{b}) The second order correlation function $g_{11,22}^{(2)}$ showing an anti-bunching dip around the single photon resonance and a bunching top around the two-photon resonance at $2\delta \approx U$. A large interaction, $U$, effectively disadvantages the doubly-occupied state, producing an effective two-level system and giving rise to strong photon anti-bunching. (\textbf{c}) Delayed $g^{(2)}$ for intermediate $U = \Gamma$ exponential and oscillating falloff with increased delay, $\tau$. (\textbf{d}) Delayed $g^{(2)}$ for strong $U = 10 \Gamma$ showing the clear asymmetry but identical oscillating period in the decay of the correlations with respect to delay time, $\tau$.}
  \label{fig:kerr}
\end{figure}

We imagine sending in photons in one channel and measuring the scattered photons in the second channel. The low order coherence functions are given by the scattering amplitudes of Eq.~\eqref{eqn:A1}-\eqref{eqn:A2}, which are calculated to be,
\begin{align}
  A^{(1)}_{1,2}  &= e^{i (\delta + 2 i \Gamma)} \frac{2 i \Gamma}{\delta + 2 i \Gamma}
  \\
  A^{(2)}_{11,22} &= 4 \Gamma^2 e^{i (\delta + 2 i \Gamma) \tau} \frac{1}{\delta +  2 i \Gamma - U/2} \frac{1}{\delta + 2 i \Gamma}
\end{align}
Allowing us to calculate the correlation functions,
\begin{align}
	g^{(1)}_{1,2} &= f e^{i \omega_0 \tau} \left|\frac{2 \Gamma}{\delta + 2 i \Gamma}\right|^2 
	\label{eqn:siteg1}
	\\
	g^{(2)}_{11,22} &=  \left| 1 - e^{i(\delta + 2 i \Gamma) \tau}\left( 1 - \frac{\delta + 2 i \Gamma}{\delta + 2 i \Gamma - U/2} \right)\right|^2
	\label{eqn:siteg2}
\end{align}
The single photon correlation function, as given by Eq.~\eqref{eqn:siteg1}, is a simple Lorentzian which allows for a single photon on resonance, $\delta = \omega_0 - \varepsilon = 0$, to perfectly transmit between the two attached channel. Figure~\ref{fig:kerr}(b) shows this first order correlation as a function of the detuning.

The $g^{(2)}$ correlation function, Eq.~\eqref{eqn:siteg2}, has a more complicated shape, as shown in Fig.~\ref{fig:kerr}(c) for different interaction strengths, $U/\Gamma$. The dip-peak shape -- which we later refer to as a Kerr signature -- shows how the photon-photon interaction shifts the two-photon resonance at $2 \delta = U$ away from the single photon resonance. At a large enough non-linear strength $U \gg \delta$ the system can effectively be modeled as a two-level emitter, showing up in the correlations through a strong anti-bunching, $g^{(2)} \ll 1$, of the transmitted photons as shown in Fig.~\ref{fig:kerr}(c) for $U=10\Gamma$.

For finite delay times between the detected photons, the two-photon correlation, $g^{(2)}(\tau)$, is plotted in Fig.~\ref{fig:kerr}(d-e) as functions of both detuning and delay time. The correlations oscillate between bunching and anti-bunching with a frequency set by the detuning, $\delta$, and a fall-off towards unity set by the total coupling strength, $2 \Gamma$.

This oscillation can be interpreted as arising from the extra phase picked up by the second photon during the delay after the first one left the scatterer. For detunings, $\delta \sim U/2$, the photons are bunched at zero delay $\tau = 0$ because the two photons scatter close to resonance, while the two photons anti-bunch at longer delay times. The multiple terms of Eq.~\eqref{eqn:siteg2} can be directly interpreted as either the elastic or the inelastic scattering amplitudes. The scattering phase shift between the inelastic an elastic contributions then readily creates the more complex oscillations patterns in the delayed correlations as shown for $U=10\Gamma$ in Fig.~\ref{fig:kerr}(d-e).
% subsection the_kerr_non_linearity (end)
\subsection{The Dimer} % (fold)
\label{sub:the_dimer}
Next, consider a Bose-Hubbard dimer consisting of two coupled Bose-Hubbard sites 
characterized by their resonance energies $\varepsilon_{1/2}$, nonlinear strengths, $U_{1/2}$, and mutual coupling strength, $t$. 

The dimer parameters and the possible contacting combinations already presents an impossibly large parameter space to explore. So far, parts of this space have been investigated numerically for both weak input power~\cite{Lee2014,Liew2010,Roy2013}, and for coherent input states of arbitrary power~\cite{Nissen2012}. In the following we show how to obtain and interpret some of these results for three different contacting geometries: parallel, perpendicular and side-coupled quasi-local.

\begin{figure*}[htb]
  \includegraphics[width=\textwidth]{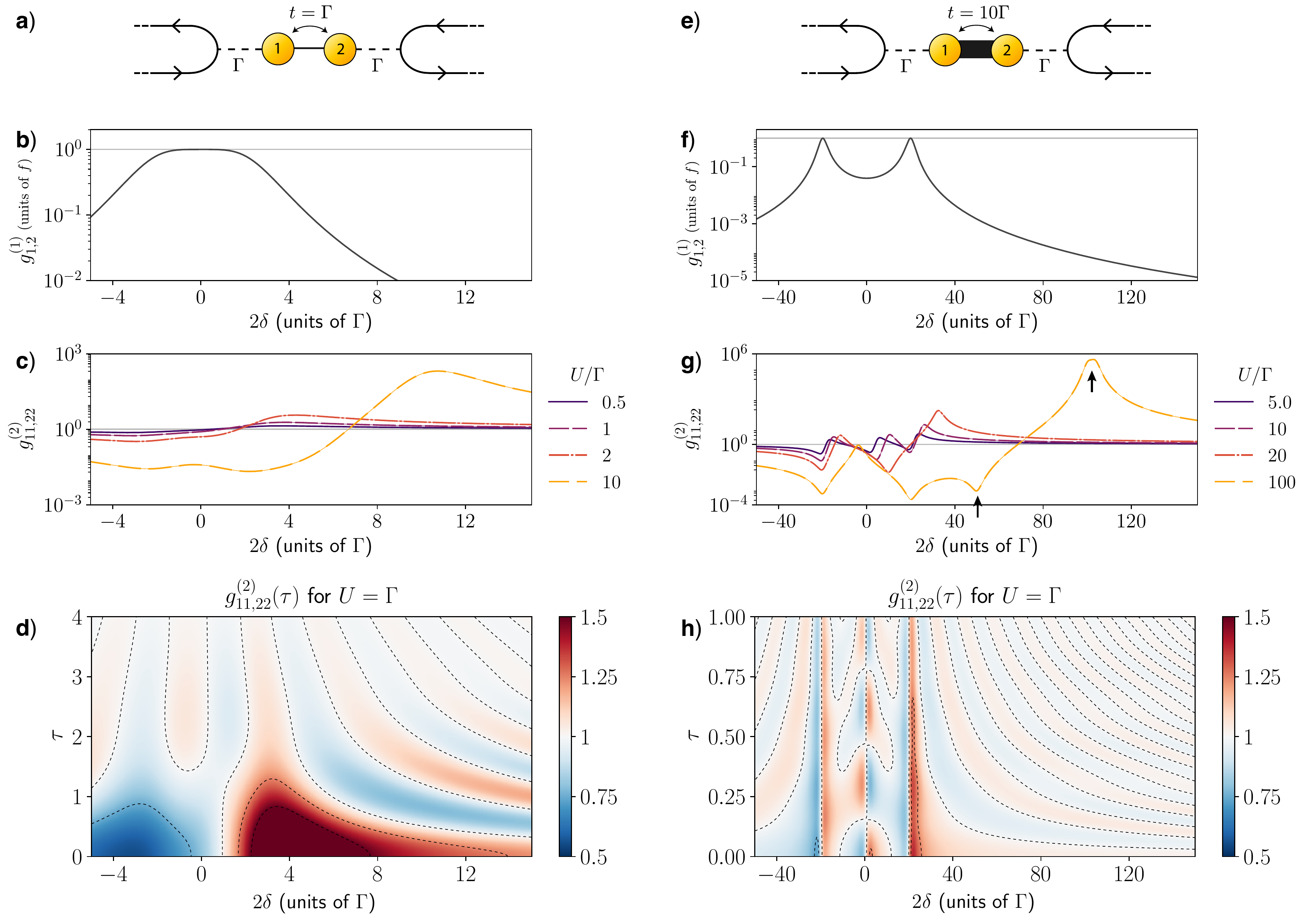}
  \caption{(Color online) Single-particle transmission and intensity-intensity correlation $g^{(2)}$ as a function of detunings for various strength of the interaction, $U$, through \textbf{(a - d)} the dimer with intermediate hopping, $t = \Gamma$, which behaves like a single non-linear cavity with an effectively smaller coupling. \textbf{(e-h)} The dimer with strong hopping, $t = 10 \Gamma$, where \textbf{(g)} shows interference features between the sharply resolved two-photon resonances, and \textbf{(h)} shows how the delayed correlation for $U = \Gamma$ distinguishes the three resonances, showing a clear difference to the single-cavity case in the oscillations around $2\delta = U = \Gamma$.}
  \label{fig:dimer}
\end{figure*}

The isolated ($\Gamma=0$) and isotropic ($\varepsilon_1 = \varepsilon_2 = \varepsilon$ and $U_1 = U_2 = U$) dimer supports two single-photon states,
\begin{align} \label{eqn:dimer1}
  |\psi^{(1)}_{l_1=\pm} \rangle &= \frac{1}{\sqrt{2}} (|1,0\rangle \pm |0,1\rangle),&& E^{(1)}_{l_1=\pm} = \varepsilon \pm t.
\end{align}
and three (here unnormalized) two-photon states:
\begin{align} \label{eqn:dimer2}
  |\psi^{(2)}_{l_2=0} \rangle &\propto |2,0\rangle - |0, 2 \rangle, \nonumber \\
  |\psi^{(2)}_{l_2=\pm} \rangle &\propto \alpha_\pm ( |2, 0\rangle + |0, 2\rangle) \pm |1, 1\rangle ,
\end{align}
with eigenenergies
\begin{align} 
E^{(2)}_{l_2=0} &= 2\varepsilon + U, \nonumber \\
E^{(2)}_{l_2=\pm} &= 2 \varepsilon + 2 \sqrt{2} t \alpha_\pm , \label{eqn:dimer_E2}
\end{align} 
given by the parameters,
\begin{align} \label{eqn:dimer2alpha}
  \alpha_\pm = \frac{1}{\sqrt{2}} \left( U/(4t) \pm \sqrt{U^2/16t^2 + 1}\right).
\end{align}
First, we couple each of the two dimer sites uniformly to their own separate chiral channel as shown in Fig.~\ref{fig:dimer}(a). For this contacting geometry, the self-energy is proportional to the total photon number, $\Sigma^{(M)} = -i M \Gamma$, which shifts the effective Hamiltonian within each photon number block, $H^{(M)} = H^{(M)}_{\text{sc}} + \Sigma^{(M)}$, the corresponding constant. The coupled and the isolated dimers then share eigenstates, $|M, l_M \rangle = |\psi^{(M)}_{l_M}\rangle$, with eigenvalues offset by the imaginary contribution from the self-energy, $\lambda^{(M)}_{l_M} = E^{(M)}_{l_M} - i M \Gamma$.

\begin{figure}[htb]
  \centering
  \includegraphics[width=\columnwidth]{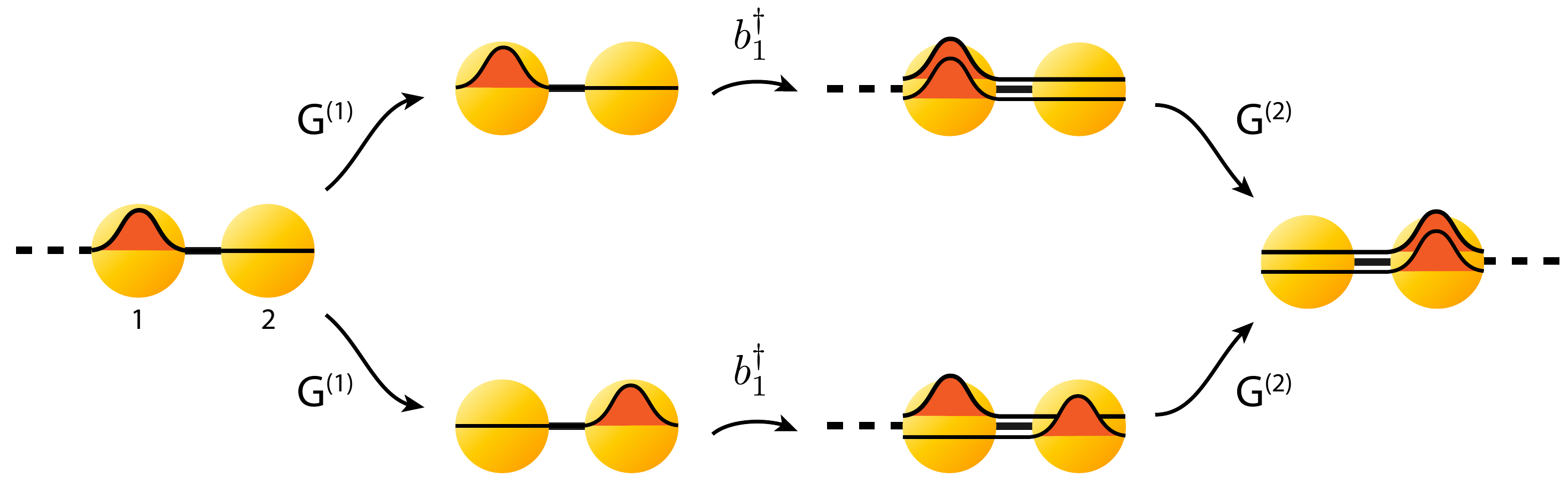}
  \caption{(Color online) Dimer interference pathways interpretation of the numerator of the last term in Eq.~\eqref{eqn:g2_2} as given by Eq.~\eqref{eqn:A2}. Starting from the left the first photon enters the dimer, and propagation of the single photon Green's function places the photon in a superposition between the two sites of the dimer. In the site basis the two dimer sites then represent the two paths which in this case gives rise to destructive interference.}
  \label{fig:dimer_interference}
\end{figure}

The uniformly coupled dimer allows for simple analytical expressions of the zero-delay amplitudes,
\begin{align}
  A^{(1)}_{1,2} &= - 2 i \Gamma \frac{t }{(\delta + i \Gamma)^2 - t^2}
  \\
  A^{(2)}_{11, 22} &= 4 \Gamma^2 
    \frac{1}{4 (\delta + i \Gamma)^2 - (\delta + i \Gamma) U - 4 t^2} 
    \nonumber \\
    & \quad\times \frac{2 t^2}{(\delta + i \Gamma)^2 - t^2 }
    \left( \frac{2(\delta + i \Gamma)}{2 (\delta + i \Gamma) - U} + 1 \right)
    \label{eqn:dimerA2}
\end{align}
with corresponding zero-delay correlation functions,
\begin{align}
	g^{(1)}_{1,2} &= f |A^{(1)}_{1,2}|^2, \\
	g^{(2)}_{11,22} &= \left|\frac{A^{(2)}_{11,22}}{(A^{(1)}_{1,2})^2} \right|^2.
\end{align}
Note that Eq.~\eqref{eqn:dimerA2} now contain two terms. It turns out these terms originate in Eq.~\eqref{eqn:A2} from two pathways the two-photon system can follow. The two pathways split off after the first photon enters the site, and either propagate to its initial site or to the other site before the second photon enters. This interpretation is also depicted graphically in Fig.~\ref{fig:dimer_interference}.

We now distinguish between dimers with intermediate and strong hopping, $t$. For intermediate hopping $t \sim \Gamma$ we plot the first and second order correlation functions in Fig.~\ref{fig:dimer}(b-d). The qualitative similarity with the single non-linear cavity in Fig.~\ref{fig:kerr} is obvious. It arises from the coupling induced level broadening for weak hopping, $t \leq \Gamma$, which does not allow the resolution of the individual dimer states. Instead the dimer behaves as a single collective level with an reduced effective coupling, $\tilde{\Gamma} = (\Gamma^2 + t^2)/(2 \Gamma) < \Gamma$, due to the total coupling being distributed over two sites instead of one. This effectively reduced coupling increases the magnitude of the correlation effects compared to that of a single non-linear Kerr element, as shown in Fig.~\ref{fig:dimer}(b) for the uniformly coupled dimer with an intermediate hopping.

For strong hopping, $t = 10 \Gamma$, the correlation functions easily resolve the individual levels as seen in Fig.~\ref{fig:dimer}(f-g). For weak interaction strengths, $U < \Gamma$, the two-photon correlation function maintains characteristic dip-peak Kerr signatures for each of the three two-photon states. 

While each of the three Kerr signatures correspond to a two-photon state, the delayed two-photon correlations clearly distinguishes the three. For weak interactions, $U \sim \Gamma$, the two side signatures at $2\delta \approx \pm 2t$ both arise from coinciding one- and two-photon states. The signature around $2\delta \approx U$ arises from a single two-photon state which does not correspond to any single-photon state. In the delayed second order correlations this difference shows up in the oscillations, since only the middle signature oscillates rapidly between bunching and anti-bunching.

The physical explanation of this oscillation is straightforward. The relevant two-photon state is the doublonic, $\ket{2,0}$, which has no components with only one photon per site. After the first photon has left the dimer, the remaining photon may perform Rabi-like oscillations between the two dimer states, thus changing the statistics at a rate proportional to the hopping between the dimer sites. For the strong hopping case ($t=10\Gamma)$ this induces oscillations of a period $T = 2\pi/10 \approx 1.59$ which fits the plot in Fig.~\ref{fig:dimer}(h).

At strong photon-photon interactions, $U \geq 4 t$, however, the $g^{(2)}$ correlations undergoes a qualitative change due to quantum interference arising from the mixing of the two-photon states as described by Eq.~\eqref{eqn:dimerA2}; this is also briefly mentioned in Ref.~\cite{Lee2014}. The last parenthesis in Eq.~\eqref{eqn:dimerA2} is the origin of the interference, which is destructive for small detunings $\delta < U$, and reaches a minimum value of $4 i \Gamma / (-U/2 + 2 i \Gamma)$ at $\delta = U/4$, thus confirming that the destructive interference becomes most pronounced in the limit of large interactions $U \gg \Gamma$, as already evident in Fig.~\ref{fig:dimer}(g). We shall see that this destructive interference in the two-photon sector is a common trait of all parallel coupled chains.

% subsection the_dimer (end)
\subsection{A Perpendicularly Coupled Dimer} % (fold)
\label{sub:a_perpendicularly_coupled_dimer}

\begin{figure}[tb]
  \centering
  \includegraphics[width=\columnwidth]{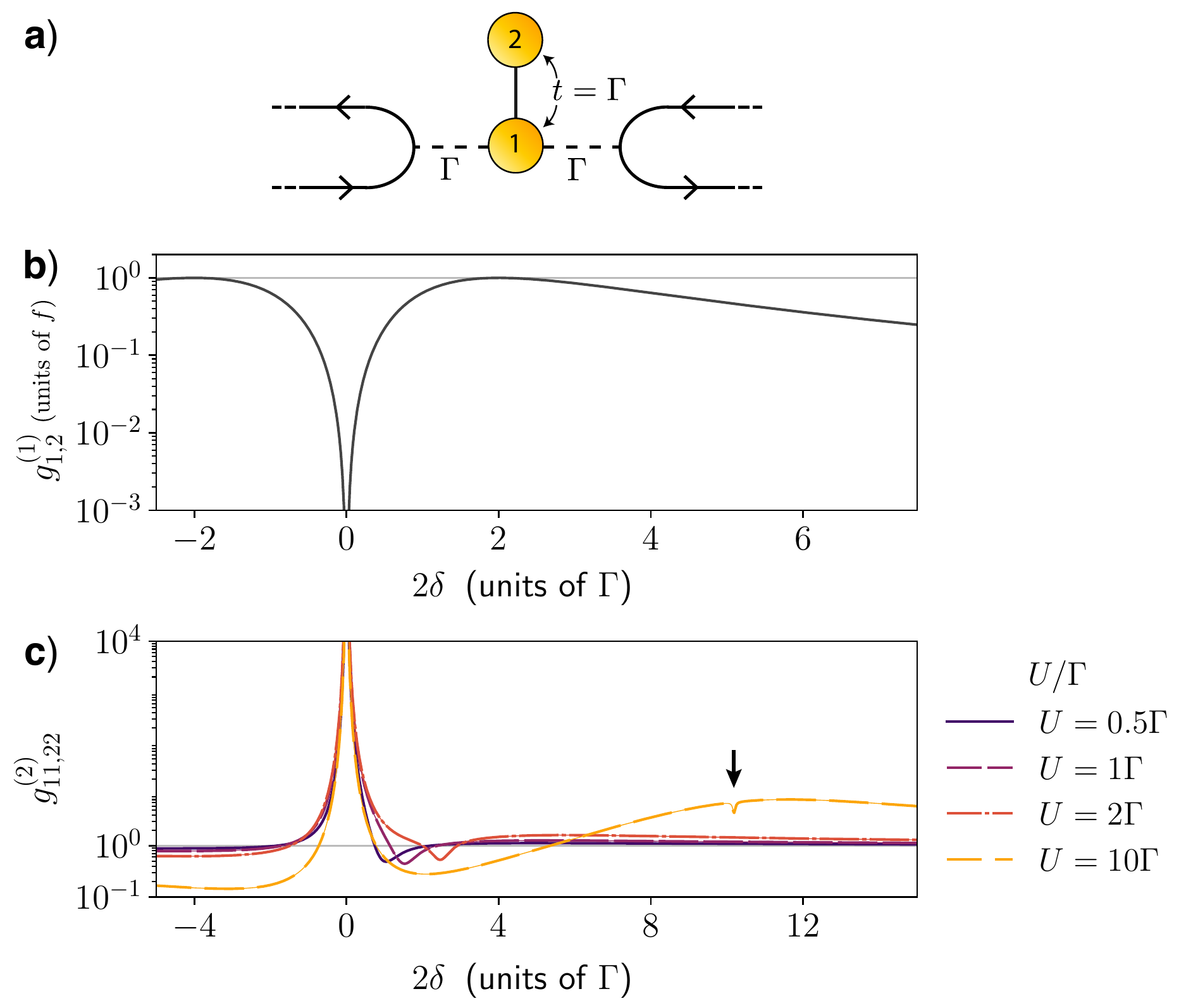}
  \caption{(Color online) The strongly coupled dimer in the side-coupled configuration. A strong destructive interference node is induced in the single-photon transmission at zero detuning as predicted by the MST rules~\cite{Markussen2010}. This induces a strong bunching effect in the $g^{(2)}$ correlations. Likewise, another interference node -- of the same geometric origin -- appears in the two-photon correlation function around $2\delta = U$ when adequately separated from the single photon resonances.}
  \label{fig:dimer_conf2}
\end{figure}

Quantum interference effects within the scatterer may render it completely transparent to single photons. A uniform Bose-Hubbard dimer connected in a perpendicular configuration, Fig.~\ref{fig:dimer_conf2}(a), constitutes a simple example of interference induced transparency. 

As shown in Fig.~\ref{fig:dimer_conf2}(b) this setup is transparent to single photons at zero detuning, meaning that the transmission between the two connected channels vanishes, and single photons are allowed to propagate in their initial channel as if uninterrupted by the scatterer.

Even slight non-linearities make the setup less transparent to two-photon states. This small discrepancy in the transparency of one- and two-photon states filters out the single-photon component of the incoming light, visible in Fig.~\ref{fig:dimer_conf2})(c) as a strong bunching peak in $g^{(2)}$ for the ``reflected'' weakly coherent light around zero detuning. For coherent input this huge bunching peak is to be ``cut off'' at a scale set by the inverse photon flux $1/f = L/(v \bar{n})$ due to higher order contributions to the few-photon expansion of $g^{(2)}$~\cite{Baranger2015}.

The transparency is perhaps unsurprising since our perpendicularly coupled dimer realizes a dressed state picture of the pumped three-level system (in a V or linear configuration). These three level systems serve as the prime example of electromagnetically induced transparency, as found experimentally by e.g.~\citet{Abdumalikov2010}. 

Transparency due to destructive single-photon interference also shows up in more elaborate lattice geometries, when the overall one-dimensionality of the scatterer and the coupling is broken, as for the perpendicularly coupled dimer. Such interference nodes can partly be predicted by the Markussen-Stadler-Thygesen (MST) rules~\cite{Markussen2010}, as we will discuss later for the ring geometry.

For strong enough non-linearities the transparency node in the single-photon transmission also shows up in the two-photon correlations. The transparency is visible in Fig.~\ref{fig:dimer_conf2} as a dip in the $g^{(2)}$ correlation around detunings $2\delta \sim U$. We can explain this behavior by looking at the generalized two-photon eigenstates. At large non-linear strengths, $U \gg t$, the three two-photon states in Eq.~\eqref{eqn:dimer2} split into the doublonic states, $\ket{0,2}$ and $\ket{2,0}$, which energetically decouple from the hardcore photonic state, $\ket{1, 1}$. The doublonic subsystem is effectively similar to the single photon system, $\ket{0,1}$ and $\ket{1,0}$, with a mutual coupling $4t/U$. This similarity carries the single-photon transparency over to the two-photon sector, and similarly induces a two-photon transparency that is offset by the doublon energy, $2\delta \sim U$. This correspondence between single photon and doublonic two-photon states is a general feature of few-photon scattering, and we will address it explicitly when discussing chain scatterers.

% subsection a_perpendicularly_coupled_dimer (end)
\subsection{A Quasi-Locally Coupled Dimer} % (fold)
\begin{figure}[tb]
  \centering
  \includegraphics[width=\columnwidth]{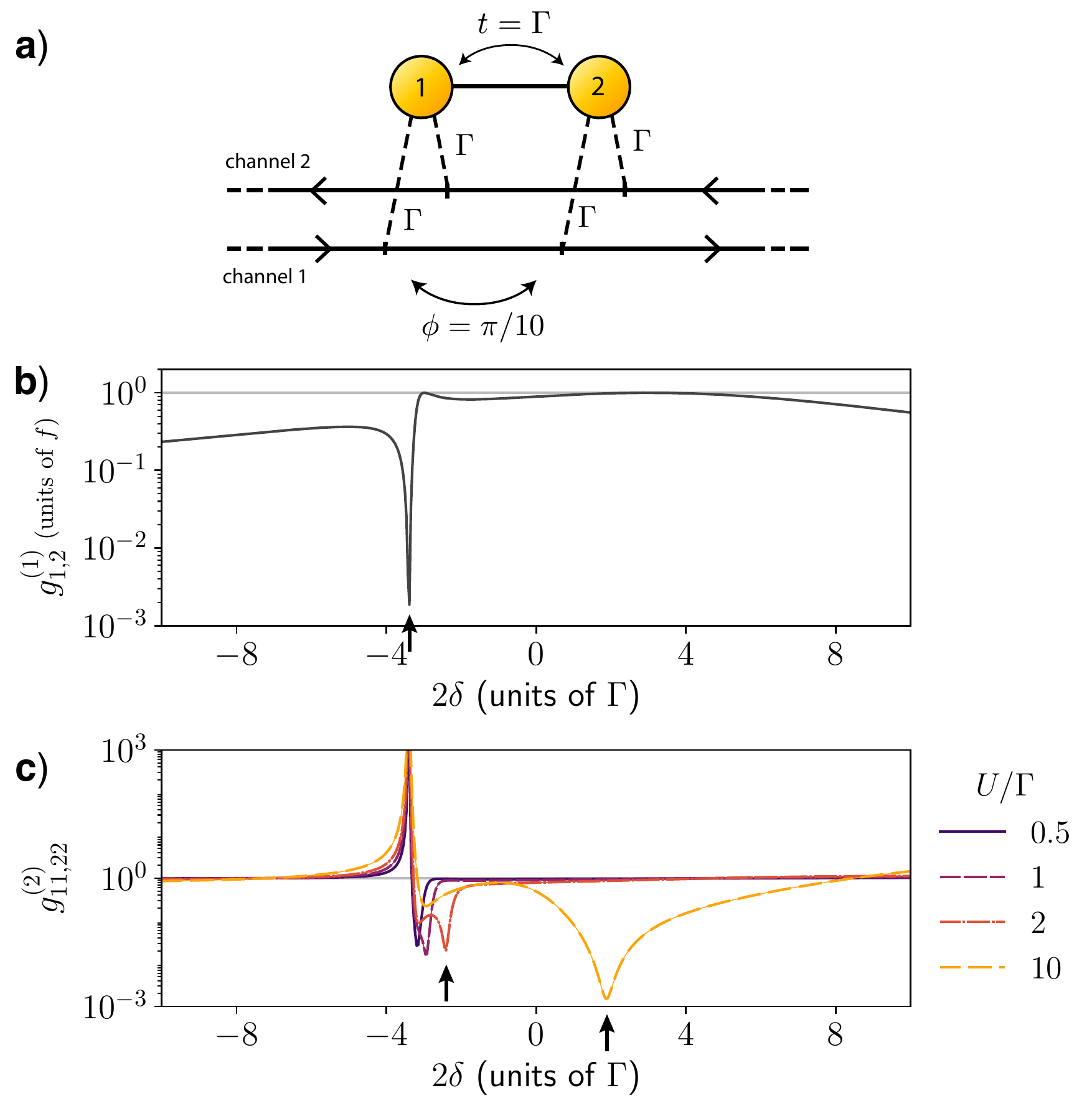}
  \caption{(Color online) \textbf{(a)} Quasi-locally coupled dimer, where each channel couples at multiple points. Within the quasi-local regime the coupling points are fully characterized by the phase difference, $\phi$. \textbf{(b)} The transmission between the two channels show a destructive interference induced transparency around $\delta = - t$}
  \label{fig:quasilocal}
\end{figure}

\label{sub:a_quasi_locally_coupled_dimer}
The dimer can also be coupled to the channels at multiple points, a setup which we can model in the quasi-local limit. We consider side-coupling the dimer symmetrically to both channels, where
\begin{align}
V &= \sum_{\sigma} \sum_{i\in\{1,2\}} (g e^{i \omega_0 x^\sigma_i} \create{a}{\sigma}\annihilate{b}{i} + h.c.),
\end{align}
The setup is shown schematically in Fig.~\ref{fig:quasilocal}(a) with the relevant phase-shift, $\phi = (x^1_2 - x^1_1) \omega_0 = (x^2_1 - x^2_2) \omega_0$, between the two coupling points.

The self-energy then becomes non-diagonal in the scatterer basis,
\begin{align}
  \Sigma^{(M)} &= -2 i \Gamma (M + \create{b}{2} \annihilate{b}{1} e^{i \phi} + \create{b}{1} \annihilate{b}{2} e^{-i \phi}).
\end{align}
This self-energy modifies the generalized eigenvalues of the dimer in a different way than for the linearly coupled dimer. For vanishing phase shifts, $\phi \approx 0$, the single-photons eigenfunctions are still given by the symmetric and anti-symmetric states of Eq.~\eqref{eqn:dimer2}, while the generalized eigenvalues become,
\begin{align}
  \lambda_\pm^{(1)} &= \ve \pm t - 2 i (1\pm1) \Gamma.
\end{align}
This means that the anti-symmetric $|\psi^{(1)}_- \rangle$ state has a vanishing imaginary part and hence decouples from the channels, making the symmetric state the only accessible single-photon state. 

When the phase shift $\phi$ is non-zero the anti-symmetric state is no longer completely isolated. The single photon transmission between channels is shown in Fig.~\ref{fig:quasilocal}(b) for $\phi = \pi/10$. For small phase differences the re-coupling to the anti-symmetric state also gives rise to destructive quantum interference and an accompanying single photon transparency.

However, in contrast to the side-coupled dimer, the two-photon correlations also develop a strong and distinct anti-bunching feature around detunings $2\delta \approx U - 2 (t + \Gamma)$ as shown in Fig.~\ref{fig:quasilocal}(c), and visible for even very small interaction strengths. While we do not go into details here, the results can be interpreted using a diagram similar to Fig.~\ref{fig:dimer_interference}.

% The two-photon states take on the same form as in Eq.~\eqref{eqn:dimer_psi2} with complex coefficients, $\alpha_\pm$, and and accompanying shifts in its eigenvalues.

{}

% subsection a_quasi_locally_coupled_dimer (end)
\subsection{The Parallel Coupled Chain} % (fold)
\label{sub:the_linearly_coupled_chain}
Longer Bose-Hubbard chains support many more single and two-photon states.  

Starting our analysis with an (isolated) uniform Bose-Hubbard chain of length, $N_s$, its single particle eigenstates are naturally indexed by their standing wave numbers $k = n \pi/ (N_s+1)$ also corresponding to the number of wave-function anti-nodes, ${n = 1, \ldots, N_s}$. The eigenenergies follow a cosine spectrum, $E_k^{(1)} = \varepsilon + 2 t \cos\left(k \right)$, and by numbering the chain sites sequentially the corresponding single-photon eigenstates can be written as,
\begin{align} \label{eqn:chain1}
  | \psi^{(1)}_k \rangle &= \sum_{i  = 1}^{N_s} \sqrt{\frac{2}{N_s +1}} \sin\left( k i \right) \create{b}{i} |0 \rangle, 
\end{align} 
We seek to understand the few photon states of the generalized Hamiltonian for uniform Bose-Hubbard chains with their first and last sites coupled to chiral channels. The generalized single photon states can be found exactly~\cite{Jin2009}, but such solutions allow for little concrete interpretation and we instead choose to analyze the one photon and two photon states perturbatively in the limits of either weak or strong coupling, $\Gamma$, combined with either a weak, $U/\Gamma \ll 1$, or a strong nonlinearity $U/\Gamma \gg 1$. We will focus on their similarity with the dimer eigenstates.

In the limit of both weak coupling and weak nonlinearity, our starting point is the eigenstates of the non-interacting chain. Here, the one-photon states are given by Eq.~\ref{eqn:chain1} and the two-photon eigenstates are the direct products of two one-photon eigenstates,
\begin{align}
  | \psi^{(2)}_{k, q} \rangle_0 &= \frac{2}{N_s+1} \sum_{i,j=1}^{N_s}  \sin(qi ) \sin(kj) \create{b}{i} \create{b}{j} |0\rangle, (k<q) \nonumber \\
   | \psi^{(2)}_{k, k} \rangle_0 &= \frac{1}{\sqrt{2}} \frac{2}{N_s+1} \sum_{i,j=1}^{N_s} \sin(ki) \sin(kj) \create{b}{i} \create{b}{j} |0\rangle.
\end{align}
At first, we focus on the zero energy subspace spanned by the $N_s /2$ two photon states (for even $N_s$) which combine two single photon states of opposite wave numbers, $|\psi^{(2)}_{k, \pi -k}\rangle_0$, $k < \frac{\pi}{2}$. The non-linearity partially lifts their zero energy degeneracy, which can be seen by writing the interaction Hamiltonian in the basis of the subspace states,
\begin{align}
  \langle \psi^{(2)}_{k,\pi-k} | H_U | \psi^{(2)}_{q, \pi - q} \rangle &=  ( 1 + \tfrac{1}{2} \delta_{k,q}) \frac{2U}{N_s+1}.
\end{align}
A single two-photon state splits off from the rest of this subspace. Like the dimer doublon state, $|2, 0\rangle$, this splinter state is also exclusively doublonic, and given by
\begin{align}
  |\phi^{(2)}_0 \rangle &= \sqrt{ \frac{2}{N_s} } \sum_{n=1}^{N_s/2} | \psi^{(2)}_{k_n, \pi-k_n} \rangle  \nonumber \\
  &= \frac{1}{\sqrt{2 N_s}} \sum_{i=1}^{N_s} (-1)^{i+1} ( \create{b}{i} )^2 \ket{0}.
\end{align}
This is actually an exact eigenstate on the isotropic Bose-Hubbard chain, with the photon-photon interaction contributing to its eigenenergy $E^{(2)}_0 =2 \varepsilon+ U$. In the perturbative analysis the rest $N_s/2 -1$ states are degenerate with an energy $2 \varepsilon+ U/(N_s+1)$ despite the finite nonlinearity.

% The $|\psi_0\rangle$ state contains doublonic states, where the two bosons inhabit the same lattice sites. This is the direct extension of the $|\psi_0\rangle$ dimer state. That the eigenenergy depends strongly on the strength of the interaction could indicate that bunched two-photon state is being carried very quickly away from the region at zero detuning. 

\begin{figure*}[htb]
  \includegraphics[width=\textwidth]{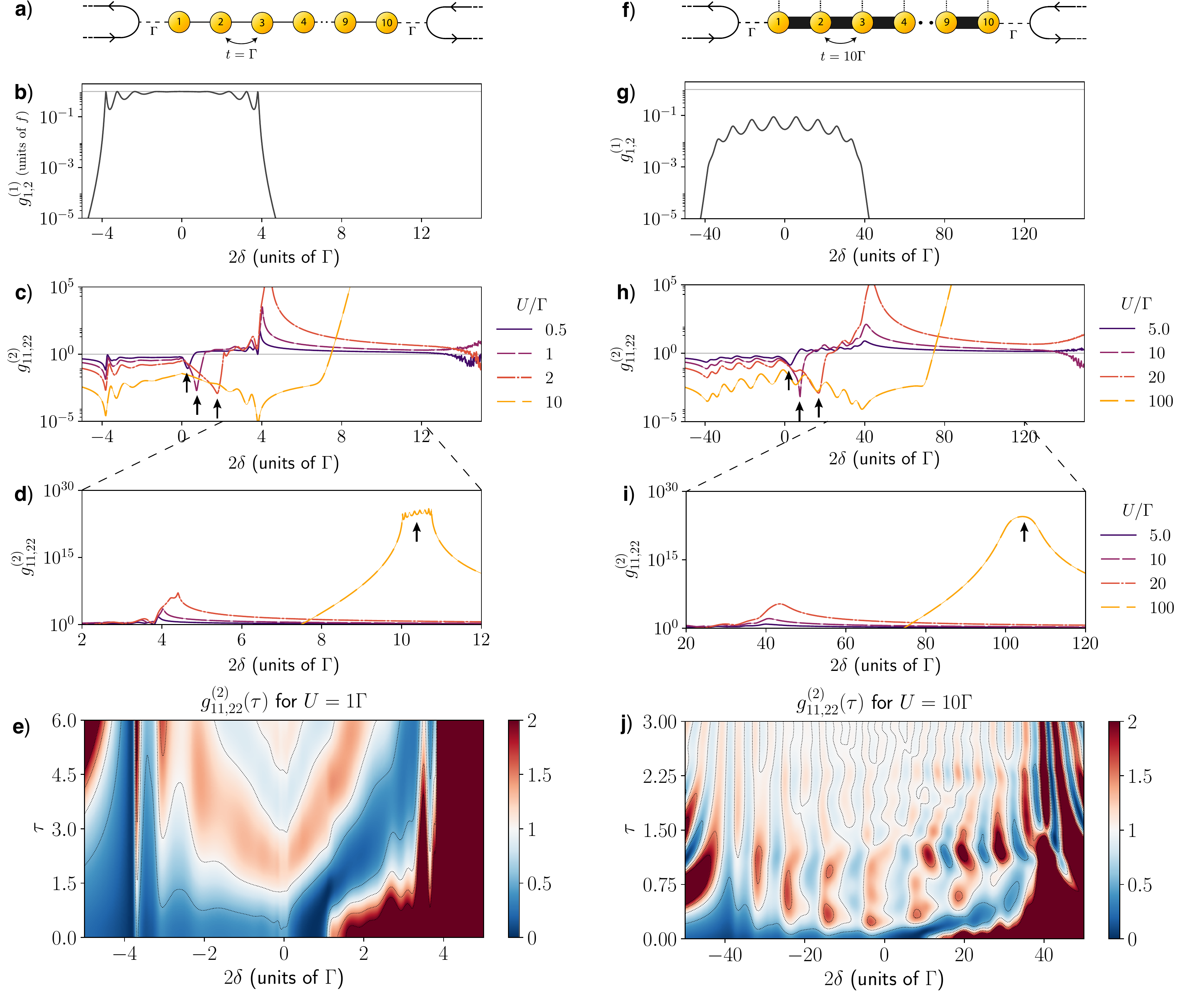}
  \caption{(Color online) Transport through 10-site chains. \textbf{(a)} The weak hopping ($t = \Gamma$) chain and corresponding \textbf{(b)} single-photon transmission and \textbf{(c-e)} second order correlation for various values of the onsite interaction strength $U$. Note the strong anti-bunching effect just below $2\delta = U$ marked with arrows. \textbf{(d)} The shape of the $g^{(2)}$ correlations for large $U$ resembles the single photon correlations as discussed in the text. Note, however, the discrepancy in the number of peaks. \textbf{(f)} The strongly coupled ($t = 10 \Gamma$) chain with additional decay channels, each with a decay rate $\Gamma_d = \Gamma/4$. \textbf{(g)} The clearly separated single-photon transmission peaks. \textbf{(h-j)} The two-photon correlation function still exhibits anti-bunching around $2 \delta = U$ marked with arrows. The delayed correlation in $\textbf{(j)}$ shows oscillations and reappearance of correlations as a function the delay time, $\tau$. }
  \label{fig:chain10}
\end{figure*}

In the regime of strong photon-photon interactions, we focus on the doublonic subspace consisting solely of states where the two photons always occupy the same site. The effective subspace Hamiltonian can be expressed through the projection, $\mathcal{P}$, onto this doublonic subspace, and the projection, $\mathcal{Q} = 1 - \mathcal{P}$, onto its complement: the subspace containing ``hard core'' photonic states. 
\begin{align}
  H^d
    &= \mathcal{P} H_U \mathcal{P} + \mathcal{P} H_t \mathcal{Q} \frac{1}{E - H_t - H_U} \mathcal{Q} H_t \mathcal{P} \\ 
    &\approx U + \mathcal{P} H_t \mathcal{Q} \frac{1}{E} \mathcal{Q} H_t \mathcal{P},
\end{align}
where in the last line we ignored contributions from additional hops beyond the initial hop which ``splits'' a doublon and the final hop which recreates it. The effective Hamiltonian for doublons on the lattice, spanned by $|2_i \rangle =\frac{(b_i^{\dagger})^2}{\sqrt{2}} | 0 \rangle$,  becomes
\begin{align}
  H^d &\approx U + \frac{2 t^2}{E} \sum_{i=1}^{N_s-1} (| 2_{i+1}\rangle \langle 2_i |+ h.c.)
    \nonumber \\
  & + \frac{2t^2}{E} \sum_{i=1}^{N_s} ( 2 - \delta_{i,1} - \delta_{i,N_s})
| 2_{i}\rangle \langle 2_i | .
\end{align}
For long chains, $N_s \gg 1$, and strong interaction, $U \gg t$, the energy difference between the two subspaces is $E \approx U$, except for a small non-uniform onsite energy at the chain ends. This roughly demonstrates that the effective doublon Hamiltonian is similar to that of a single photon on the same chain with a slight discrepancy in the on-site energies at the chain ends. The two-photon eigenenergies in this subspace then form a cosine spectrum ${E^d_k = 2 \varepsilon+ U  + 4t^2/U (1 + \cos(k))}$, and the eigenstates are here standing waves of doublons, ${|\psi^d_k \rangle = \sqrt{2/(N_s+1)} \sum_{i = 1}^{N_s} \sin(k i) |2_i \rangle}$.

In the limit of an infinitely strong photon-photon repulsion, $U/t \rightarrow \infty$, the remaining $\frac{N_s (N_s-1)}{2}$ two-photon states describe hard core photons forming the lattice equivalent of a Tonks-Girardeaux gas~\cite{Cazalilla2011}. Their eigenfunctions are analogous to those of non-interacting fermions, but with bosonic statistics,
\begin{align}
  |\psi^{TG}_{k , q} \rangle = \sum_{i, j} S(i, j) \langle i | \psi_k^{(1)} \rangle \langle j | \psi_q^{(1)} \rangle \create{b}{i} \create{b}{j} | 0 \rangle, \quad k<q .
\end{align}
The fermionized behavior is captured by the exchange sign $S(i, j) = 2 \Theta(i-j) - 1$, with $\Theta$ being the Heaviside step function. Obviously, $S (i,i)=0$. These states play the same role as the $|\psi_{l_2 = \pm}^{(2)} \rangle$ dimer eigenstates from Eq.~\eqref{eqn:dimer2}.

This picture becomes more complicated when we wish to include the self-energy arising from the coupling to the chiral channels as discussed in Eq.~\eqref{egn:Sigmasigma}. Starting from the non-interacting limit, we may consider the case of a perturbatively small coupling $\Gamma$ (compared to the relevant energy scales). To lowest order the eigenenergies are modified through the wavefunction density at the coupling sites (located at the chain ends),
\begin{align}
  E^{(1)}_k - \varepsilon \approx 2 t \cos(k) - \frac{4}{N_s+1} i \Gamma \sin^2(k).
\end{align}

In the strong coupling (weak hopping) limit one may instead start from the subsystem that excludes the coupling sites. The subsystem quasi-momenta are then given by $k = \pi m /(N_s-1)$, with integer $m = 1, \ldots, N_s-2$ and eigen-energies $E^{(1)}_k - \varepsilon = 2 t \cos(k)$. The system also supports two additional eigenstates localized around the coupling sites and with imaginary eigen-energies approximately given by $E^{(1)}_{\sigma} - \varepsilon = -i \Gamma_\sigma$.

Figure~\ref{fig:chain10} shows the few-photon correlation function for 10-site chains with intermediate hopping $t=\Gamma$ and strong hopping $t=10\Gamma$. In the case of strong hopping, we also attach additional decay channels to each of the chain sites with a decay strength, $\Gamma_d = \Gamma/4$, in order to both demonstrate the capability of the scattering method to deal with such systems, and to smooth out the otherwise sharp features of the correlation functions.

The two-photon correlations for both chains develops a strong anti-bunching feature around $2\delta = U$ present for even small values of the photon-photon interaction. We speculate that this feature has the same origin as the anti-bunching feature for the parallel coupled dimer, and is connected to the exclusively doublonic state $|\phi_0^{(2)}\rangle$ with the energy $E=U$. It shows up for chains of all lengths, but is most pronounced for longer chains. We also note the robustness of this feature, as it is present even for a chain with strong hopping and additional decays.

The similarity between the doublonic subspace at large values of the photon-photon interaction strengths and the single photon Hilbert space is visible in Fig.~\ref{fig:chain10}(d) and (i). The band-width of the doublonic spectrum in Fig.~\ref{fig:chain10}(d) is here roughly $8t^2/U = 0.8$ for the case of intermediate hopping, $t=\Gamma$. However, we already discussed that the coupling at the ends, means that the end sites do not contribute in the same way to the Hilbert space, and we may partly exclude them when calculating the Green's function; and when counting, the two-photon correlation function only shows eight resonance peaks as opposed to the ten resonance peaks in the single-photon correlation function.

For large hopping the additional decays smears our the shape of the correlations as in Fig.~\ref{fig:chain10}(i).

The delayed two-photon correlation functions are shown in Fig.~\ref{fig:chain10}(e) and (j) for $U=\Gamma$ and $U=10\Gamma$ respectively. Both show the decay of the anti-bunching feature around $2 \delta = U$, and both show oscillations on a time-scale set by $U + |2\delta|$. For large hoppings the oscillations only lasts for one or two periods before decaying through the additional channels. However, the oscillations here are reminiscent of revivals with strong correlations suddenly emerging at finite delay times from previously weak values. While reminding us of previous results~\cite{Laakso2014}, we speculate that one may solve the full non-Markovian dynamics by simulating a one-dimensional waveguide  using scattering theory on a discrete chain. A notion one may pursue in further work.

% Since the effective doublon hopping amplitude is inversely proportional to the onsite photon repulsion, the doublons can always be driven into the strongly coupled (weak hopping) regime by a large enough nonlinear term, $4t^2/U < 2 \Gamma$. While the localized end states couple strongly to the channels, the associated bunching effect vanishes on a timescale $1/2\Gamma$. The localized states isolates the bulk states from the transmission channel, and lowers their effective coupling to the chiral channels (to lowest order in perturbation theory, ${\Sigma_d \propto - i 4 t^2 \Gamma/U^2}$). The effective weak coupling reduces the transmission of the two-photon states, but allows the bunching to persist for much longer time. When tuned outside the single-photon band, defined by detunings,
% \begin{align}
% 2\delta \in [-2 t \cos(\pi/(N_s+1)), 2t \cos(\pi/(N_s+1))]
% \end{align}
% the doublonic system delivers very large values of the two-photon correlation, $g^{(2)}$, corresponding to strong bunching even for large spatial (or temporal) photon separation.

% subsection the_linearly_coupled_chain (end)
\subsection{Rings} % (fold)
\begin{figure}[htb]
  \centering
  \includegraphics[width=\columnwidth]{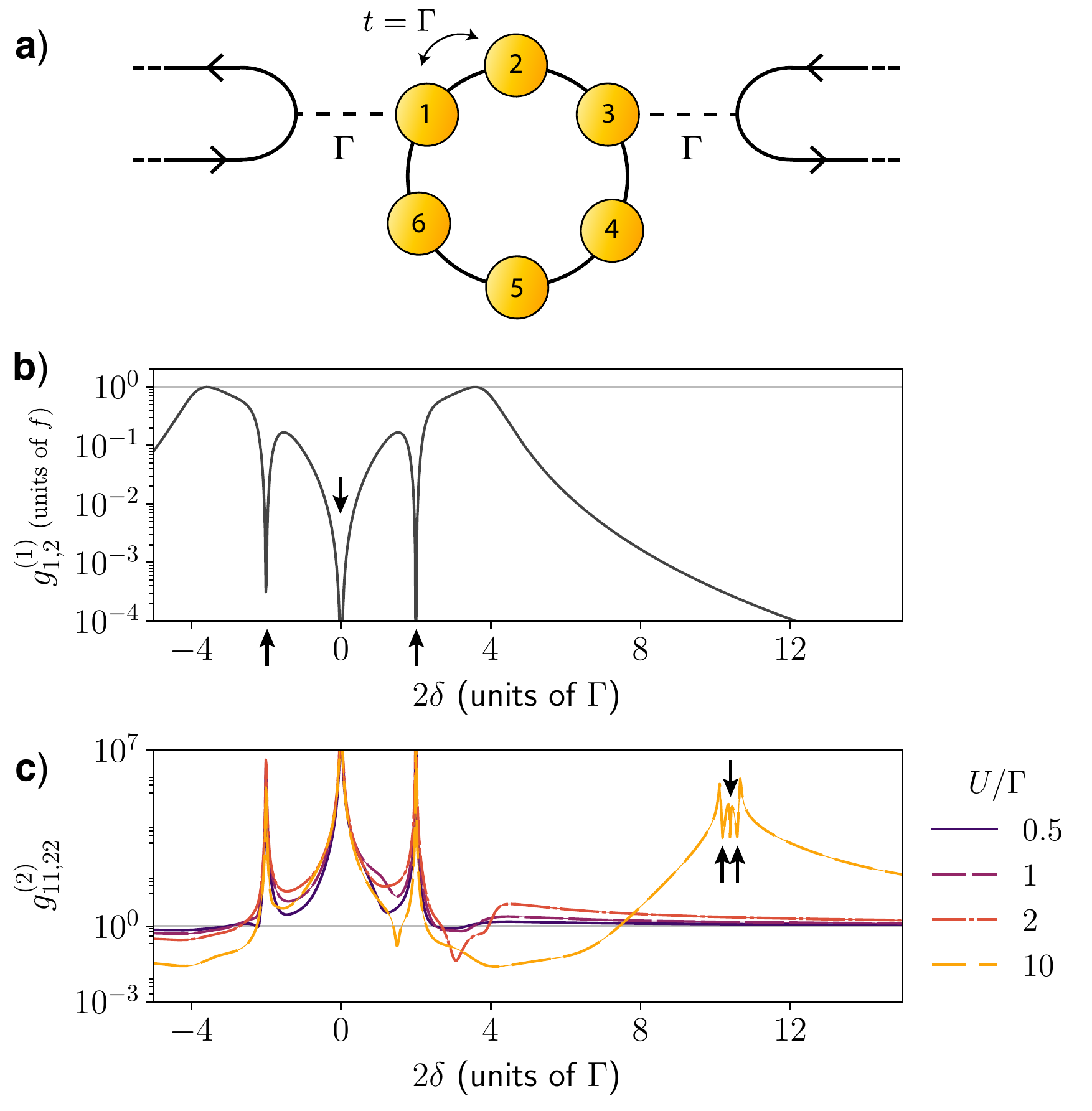}
  \caption{(Color online) Transmission and second order intensity correlation for the six site ring lattice with weak hopping, $t=1\Gamma$. \textbf{(a)} The single-photon transmission with the characteristic three dip structure arising as a consequence of quantum transport interference. \textbf{(b)} The intensity-intensity correlation function as a function of detuning for various values of the interaction, $U$. For strong repulsion, $U \geq 4\Gamma$, the three dip structure of the single-photon transmission is replicated in the intensity-intensity correlation due dominance of the (mostly) doublonic subspace at those detunings as annotated with arrows.}
  \label{fig:ring}
\end{figure}

\label{sub:rings}
Next, we consider six sites arranged in a ring geometry with two chiral channels coupling to two different sites on the ring as shown in Fig.~\ref{fig:ring}(a).
% The isolated ring support single photon states Bloch states characterized by a wavenumber $k=2 m\pi/N_s$ for $m = 0, \ldots, N_s-1$, with energy $E^{(1)}_k = 2 t \cos(k)$, and corresponding wavefunctions,
% \begin{align}
%   % |\psi^{(1)}_k \rangle =\sqrt{\frac{2}{N_s}}  \sum_{i=1}^{N_s} \cos(k i) \create{b}{i} \ket{0} , \sqrt{\frac{2}{N_s}}  \sum_{i=1}^{N_s} \sin(k i) \create{b}{i} \ket{0} \\
%   k=0,\ldots , \frac{N_s}{2} , \quad k=1, \ldots ,\frac{N_s}{2}-1 (N_s \,\, \text{even}) \\
%   k=0,\ldots, \frac{N_s-1}{2}, \quad k=1,\ldots, \frac{N_s-1}{2} (N_s \,\, \text{odd}).
% \end{align}
% [I ADDED MORE EXPLICIT FORMULAS HERE. DO WE NEED THEM?]
By coupling channels at different sites, one may induce destructive quantum interference effects which block single-photon transport. This is clearly visible in Fig.~\ref{fig:ring}(b) where the transmission through the six-sited ring shows three interference nodes. As already discussed in our analysis of the perpendicularly coupled dimer, this effect is geometrical and can be found in many other lattice geometries where the one-dimensionality of the combined channel-lattice system has been broken.

% Rings with an even number of sites, has a bipartite lattice with an equal number of lattice sites in either sublattice. For istropic bipartite lattices in general, coupling to two sites within the same sublattice~\fixme{cite} blocks the single photon transmission at zero detuning in accordance with the MKS rules~\cite{Markussen2010}. 
The blocking of single photon transport around the interference nodes again induces a large relative bunching effect as seen in the plot of the two-photon correlation in Figure~\ref{fig:ring}(c), because the finite photon-photon interaction lifts the transport blockade and allows the two-photon states to transmit.

The perturbative analysis for weak non-linearity performed for the chain geometry easily carries over to the ring geometry: Again a single two-photon state consisting entirely of doublons, factors out, and creates a deficit of bunched states around zero detuning visible in the $g^{(2)}$ function as an anti-bunching dip. 

In the regime of strong non-linearity, the states separate again into two subspaces describing doublons and hardcore photons. Figure.~\ref{fig:ring} shows that the characteristic three node pattern int the single photon transmission is now also visible in the doublonic part of the two-photon correlation at detunings $2\delta \approx U$.

% subsection rings (end)
\subsection{Planes and Edge States} % (fold)
\label{sub:planes_and_edge_states}

A Bose-Hubbard lattice plane penetrated by a magnetic field is known to support single-particle topological quantum Hall edge states~\cite{Hofstadter1976} as has also been investigated experimentally~\cite{Hafezi2013a}. 

\begin{figure}[htb]
  \centering
  \includegraphics[width=\columnwidth]{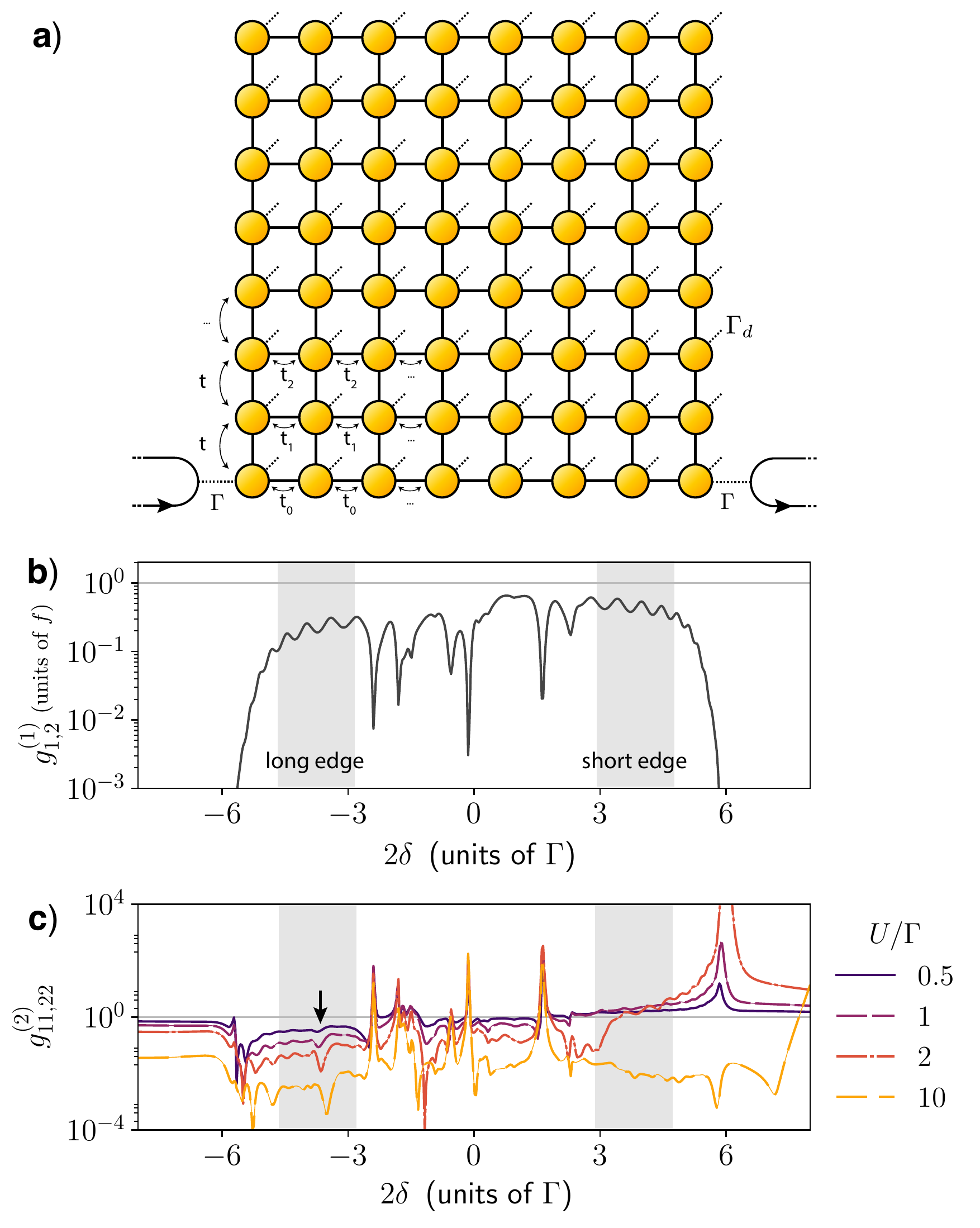}
  \caption{(Color online) Transmission and $g^{(2)}$ function for a plane penetrated by a (artificial) magnetic field. \textbf{(b)} The single photon transmission amplitude between the attached channels. Such systems support single-photon edge states that in our case follow the lattice edge either the long or the short way around. The parts of the spectrum where edge states dominate have been shaded gray. \textbf{(c)} The $g^{(2)}$ correlation for transport along the long edge clearly resembles the correlations of a simple chain with the characteristic anti-bunching dip, thus making the one-dimensionality of the edge state clearly discernible (compare Fig.~\ref{fig:chain10}).}
  \label{fig:plane}
\end{figure}

We consider a uniform 8 by 8 square lattice of Bose-Hubbard sites shown schematically in Fig.~\ref{fig:plane}, where the hopping Hamiltonian takes the form
\begin{align}
  H_{t} &= \sum_{\langle i,j \rangle} \left(t_{i,j} \create{b}{i} \annihilate{b}{j} + h.c.\right),
\end{align}
with hopping amplitudes
\begin{align}
  t_{i,j} = \begin{cases}
    t & \text{ if } i \text{ and } j \text{ in different rows.} \\
    t_k = t e^{ik\phi} & \text{ if } i \text{ and } j \text { oth in row } k.
  \end{cases}
\end{align}
Such a row-increasing phase has already been artificially induced in optical resonator circuits~\cite{Hafezi2013a}, and can be interpreted as an effective magnetic field. The system should then exhibit the quantum Hall effect, and its single particle states should separate into either edge or bulk states.

In order to make the setup more experimentally realistic we added an additional decay channel to each of the Bose-Hubbard sites--although with a rather small coupling rate, $\Gamma_d = \Gamma/100$.

We finally attach chiral channels to two lattice corners at the usual coupling strength, $\Gamma = t$. For an effective phase per plaquette of $\phi = 2\pi/5$, the resulting first and second order correlation function is shown in Fig.~\ref{fig:plane}. We have explicitly marked the part of the spectrum containing edge states in the the single-photon transmission in Fig.~\ref{fig:plane}(b). The absence of interference nodes in those parts of the spectrum makes the qualitative resemblance to the transmission through a linear chain in Fig.~\ref{fig:chain10} obvious. The bulk spectrum on the other hand shows several transmission nodes at irregular values of the detuning.

The edge states at the two ends of the total spectrum propagate in opposite directions, with the edge states in the low end of the spectrum traveling the long way around the lattice, while the high end edge states travel the short way around.

The $g^{(2)}$ correlation for transport along the long edge clearly resembles the correlations of a simple chain with the distinct anti-bunching dip in the middle of the edge state spectrum clearly visible, thus further demonstrating the one-dimensionality of the chiral edge states.
% subsection planes_and_edge_states (end)
% section results (end)
\section{Conclusion} % (fold)
\label{sec:conclusion}
In conclusion, we have investigated few photon scattering in systems with many internal degrees of freedom. Specifically we have focused on systems with multiple chiral channels coupling to extended Bose-Hubbard lattices. In the case of local couplings -- coupling to each channel at a single local point --  we have derived exact analytical expressions for the single-photon and two-photon scattering matrices. In the case of multiple coupling points we have derived similar expressions for the single-photon and two-photon scattering matrices in the quasi-local regime, which we have argued is equivalent to invoking the Markov approximation.

We have implemented the analytical expression numerically and shown results for the scattering of weakly coherent light on several different Bose-Hubbard graphs. 

Obviously, we have by no means exhausted the possibilities for analyzing photon scattering on Bose-Hubbard graphs. We therefore offer babusca -- a Python implementation of the results obtained in this paper -- for download online~\footnote{The latest babusca library is available at \href{http://github.com/georglind/babusca}.} and as an attachment to the supplementary material for this paper. Babusca makes it straightforward to analyze the low-order coherence functions for photon transport through chiral channels that couple locally or quasi-locally to an arbitrary Bose-Hubbard lattice.

During the finishing stages of this manuscript, we became aware of similar work carried out by~\citet{See2017}, and we refer to their manuscript for a complementary but equivalent formulation of few photon scattering on Bose-Hubbard lattices.

% section conclusion (end)
\clearpage
\appendix

\newcommand{\der}[2]{{\frac{d #1}{d #2}}}
\def\bgrk#1{\mbox{{\boldmath $#1$ \unboldmath}}\!\!}
\def\eq#1{(\ref{#1})}
\def\H1{\widehat{H}_1}
\def\M{\hbox{\large \tt M}}
\def\dmat{\hbox{\large \tt d}}
\def\G{\slash\mkern-11muG}
\def\ug{\underline{\gamma}_2}
\def\pl{\partial_{\Lambda}}
\renewcommand{\i}{\ensuremath{\mathrm{i}}}
\newcommand{\e}{\ensuremath{\mathrm{e}}}
\renewcommand{\d}{\ensuremath{\mathrm{d}}}
\newcommand{\pd}{\partial}

\section{Two photon scattering} \label{app:local}

The two-photon $T$ matrix is given by~\eqref{eqn:TN}. In explicit form it reads
\begin{widetext}
\begin{align}
T^{(2)} &= P^{(0)} \, \vvdots \, V G V G V G V \, \vvdots \, P^{(0)}
\nonumber \\
&= P^{(0)} \, \vvdots \, \create{a}{\nu'_1} \annihilate{\tilde{b}}{\sigma'_1} 
G(E)
\left[\create{a}{\nu'_2} \annihilate{\tilde{b}}{\sigma'_2} + \create{\tilde{b}}{\sigma_2} \annihilate{a}{\nu_2} \right] 
G(E) 
\left[\create{a}{\nu'_2} \annihilate{\tilde{b}}{\sigma'_2}  + \create{\tilde{b}}{\sigma_2} \annihilate{a}{\nu_2}\right]
 G(E) 
 \create{\tilde{b}}{\sigma_1} \annihilate{a}{\nu_1} \,\vvdots\, P^{(0)}
\nonumber \\
&\stackrel{\text{o.s.}}{=}
% on shell G2 part
P^{(0)} \annihilate{\tilde{b}}{\sigma'_1} G^{(1)}(E + \omega'_1+ \omega_0)  \annihilate{\tilde{b}}{\sigma'_2}  G^{(2)}(E+ \omega'_1+ \omega'_2+2 \omega_0) \create{\tilde{b}}{\sigma_2} G^{(1)} (E + \omega_1+ \omega_0) \create{\tilde{b}}{\sigma_1} P^{(0)} \create{a}{\nu'_1} \create{a}{\nu'_2}  \annihilate{a}{\nu_2} \annihilate{a}{\nu_1}
\nonumber \\
% on shell G0 part
&\qquad +  P^{(0)}  \annihilate{\tilde{b}}{\sigma'_1}  G^{(1)} (E + \omega'_1+ \omega_0) \create{\tilde{b}}{\sigma_2} G^{(0)} (E + \omega'_1 -\omega_2)
\annihilate{\tilde{b}}{\sigma'_2} G^{(1)}(E + \omega_1 + \omega_0) \create{\tilde{b}}{\sigma_1} P^{(0)}  \create{a}{\nu'_1} \create{a}{\nu'_2} \annihilate{a}{\nu_2} \annihilate{a}{\nu_1} 
\nonumber \\
&= P^{(0)}  \annihilate{\tilde{b}}{\sigma'_1} \sum_{l'_1} \frac{P_{l'_1}^{(1)}}{\omega'_1 + \omega_0 - \lambda_{l'_1}^{(1)}}  \annihilate{\tilde{b}}{\sigma'_2} \sum_{l_2} \frac{P_{l_2}^{(2)}}{\omega'_1+ \omega'_2+2 \omega_0 - \lambda_{l_2}^{(2)}} \create{\tilde{b}}{\sigma_2} \sum_{l_1} \frac{P_{l_1}^{(1)}}{\omega_1 + \omega_0 - \lambda_{l_1}^{(1)}} \create{\tilde{b}}{\sigma_1} P^{(0)} \create{a}{\nu'_1} \create{a}{\nu'_2} \annihilate{a}{\nu_2} \annihilate{a}{\nu_1}
\nonumber \\
&\quad +  P^{(0)}  \annihilate{\tilde{b}}{\sigma'_1} \sum_{l'_1}  \frac{P_{l'_1}^{(1)}}{\omega'_1+ \omega_0 - \lambda_{l'_1}^{(1)}} \create{\tilde{b}}{\sigma_2} \frac{P^{(0)}}{\omega'_1 - \omega_2 + i 0^+}
 \annihilate{\tilde{b}}{\sigma'_2}  \sum_{l_1}  \frac{P_{l_1}^{(1)}}{\omega_1 + \omega_0 - \lambda_{l_1}^{(1)}} \create{\tilde{b}}{\sigma_1} P^{(0)} \create{a}{\nu'_1} \create{a}{\nu'_2} \annihilate{a}{\nu_2} \annihilate{a}{\nu_1}.
\label{eqn:t2_BH}
\end{align}
\end{widetext}
Note the implicit summation over all four channel state indices, $\nu'_1, \nu'_2, \nu_1, \nu_2$. From the third line on we make use of the on-shell (o.s.) condition $E= \omega'_1 + \omega'_2 = \omega_1 + \omega_2$. In the last two lines we set $E-H_{\text{chs}} =E-( \omega'_1 + \omega'_2) =0$, since the operator  $E-H_{\text{chs}}$ after the (modified) normal ordering acts on the final state. 

In the following we consider only the principal value part of $(\omega'_1 - \omega_2 + i 0^+)^{-1}= (\omega'_1 - \omega_2)^{-1} - i \pi \delta(\omega'_1 - \omega_2)$ in \eqref{eqn:t2_BH}, because the delta function part contributes to the elastic two-photon scattering accounted by the term $\propto S^{(1)} S^{(1)}$ in \eqref{eqn:S2}. The spurious divergence $\omega'_1 = \omega_2$ in $(\omega'_1 - \omega_2)^{-1}$ is removed after symmetrization of the last line term in \eqref{eqn:t2_BH} under permutation of the field indices $\nu'_1$ and $\nu'_2$. Thus we obtain
\begin{widetext}
\begin{align}
T^{(2) \mathcal{P}}
  &= \frac{P^{(0)}}{2}\create{a}{\nu'_1} \create{a}{\nu'_2} \annihilate{a}{\nu_2} \annihilate{a}{\nu_1} \sum_{l'_1, l_1} \left\{ 2 \sum_{l_2} \frac{ \langle 0| \annihilate{\tilde{b}}{\sigma'_1} P_{l'_1}^{(1)}\annihilate{\tilde{b}}{\sigma'_2} P_{l_2}^{(2)} \create{\tilde{b}}{\sigma_2}  P_{l_1}^{(1)} \create{\tilde{b}}{\sigma_1} | 0 \rangle}{(\omega'_1 + \omega_0 - \lambda_{l'_1}^{(1)}) (\omega_1 + \omega_2 +2  \omega_0- \lambda_{l_2}^{(2)}) (\omega_1 + \omega_0-\lambda_{l_1}^{(1)})}  
    \right.
  \nonumber \\
  & \left. \qquad \qquad \qquad \qquad  \!\!+ \frac{\langle 0 | \annihilate{\tilde{b}}{\sigma'_1}  P_{l'_1}^{(1)} \create{\tilde{b}}{\sigma_2} | 0 \rangle \langle 0 |\annihilate{\tilde{b}}{\sigma'_2} P_{l_1}^{(1)} \create{\tilde{b}}{\sigma_1} | 0 \rangle}{(\omega'_1 + \omega_0  - \lambda_{l'_1}^{(1)}) (\omega'_1 - \omega_2) (\omega_1 + \omega_0 -\lambda_{l_1}^{(1)})}
    - \frac{\langle 0 | \annihilate{\tilde{b}}{\sigma'_2}  P_{l_1}^{(1)} \create{\tilde{b}}{\sigma_1} | 0 \rangle \langle 0 | \annihilate{\tilde{b}}{\sigma'_1}   P_{l'_1}^{(1)} \create{\tilde{b}}{\sigma_2} | 0 \rangle}{(\omega'_2 + \omega_0 - \lambda_{l_1}^{(1)}) (\omega'_1 - \omega_2) (\omega_2 + \omega_0 -\lambda_{l'_1}^{(1)})}  \right\}.
\label{t2_BHb}
\end{align}
Combined with the identity
\begin{align*}
& \frac{1}{(\omega'_1 + \omega_0- \lambda_{l'_1}^{(1)}) (\omega_1 + \omega_0 -\lambda_{l_1}^{(1)})}   - \frac{1}{(\omega'_2 + \omega_0 - \lambda_{l_1}^{(1)}) (\omega_2 + \omega_0 -\lambda_{l'_1}^{(1)})} \\
& \quad \stackrel{\text{o.s.}}{=} - \frac{(\omega'_1 - \omega_2)(E - \lambda_{l'_1}^{(1)} -\lambda_{l_1}^{(1)})}{(\omega'_1 + \omega_0 - \lambda_{l'_1}^{(1)}) (\omega'_2 + \omega_0 - \lambda_{l_1}^{(1)}) (\omega_2 + \omega_0 - \lambda_{l'_1}^{(1)}) (\omega_1 + \omega_0 - \lambda_{l_1}^{(1)})},
\end{align*}
this directly produces Eq.~\eqref{eqn:T2}.
\end{widetext}

\section{Second-order intensity correlation function} \label{app:g2}
\label{sec:Second_order_intensity_correlation}

This sections details the derivation of the second order intensity correlation function, $g^{(2)}$, for weakly coherent light. The lowest order contribution to $g^{(2)}$ involves the two photon component of the coherent state, ${|\phi_{\nu_0}^{(2)} \rangle = \tfrac{1}{2} \bar{n} (\create{\mathcal{A}}{\nu_0 } (t))^2 \ket{0,0}}$, which scatters into the state ${| S \phi^{(2)}_{\nu_0} \rangle = S^{(2)} | \phi_{\nu_0}^{(2)} \rangle}$. The latter is constructed using Eqs.~\eqref{eqn:S2} and \eqref{eqn:T2}, resulting in 
\begin{widetext}
\begin{align}
    | S^{(2)} \phi^{(2)}_{\nu_0} \rangle &= \bar{n} e^{2 i \omega_0 t}
      \left(
       \frac12 s^{(1)}_{\sigma'_2 \sigma_0 }  s^{(1)}_{\sigma'_1 \sigma_0 }  \create{\mathcal{A}}{\sigma'_2 \omega_0} \create{\mathcal{A}}{\sigma'_1 \omega_0}
      - 2 \pi i T^{(2) \mathcal{P}}_{\sigma'_1 \sigma'_2; \sigma_0 \sigma_0} (\omega'_1, \omega'_2; 0,0) \frac{2 \pi}{L} \create{a}{\sigma'_2 \omega'_2} \create{a}{\sigma'_1 \omega'_1}\, \delta_{\omega'_1 + \omega'_2,0} \right) |0 ,0 \rangle ,
  \end{align}
with implicit summations over repeated indices.

Next step involves the evaluation 
\begin{align}
  &
  \annihilate{a}{\sigma} (x-t-\tau) \annihilate{a}{\sigma'} (x -t) | S^{(2)} \phi^{(2)}_{\nu_0} \rangle \nonumber \\
  &\quad = 
    \frac{\bar{n}}{L} e^{i \omega_0 (2 x +  \tau) } \left[ s^{(1)}_{\sigma'  \sigma_0} s^{(1)}_{\sigma \sigma_0} - 2 \pi i
      \int_{-\infty}^{\infty} \id{\omega} (\delta_{\sigma, \sigma'_1} \delta_{\sigma', \sigma'_2} e^{-i \omega \tau} + \delta_{\sigma, \sigma'_2} \delta_{\sigma', \sigma'_1} e^{i \omega \tau} ) T^{(2) \mathcal{P}} _{\sigma'_1 \sigma'_2; \sigma_0  \sigma_0} (\omega, - \omega; 0,0) \right] \ket{0,0} .
\end{align}
The $\omega$ integral can be performed by closing the contour of integration in either the upper or lower complex half-plane depending on the exponential prefactor (also recall that $\mathrm{Im} \lambda_{l_M}^{(M)} < 0 $). After the proper normalization we obtain
\begin{align} \label{eqn:g2_app_res}
  g_{\sigma \sigma'}^{(2)} (\tau) 
  &=
    \frac{
    	\left|\left| \annihilate{a}{\sigma} (x- t - \tau) \annihilate{a}{\sigma'} (x -t) | S^{(2)} \phi^{(2)}_{\nu_0} \rangle \right|\right|^2}
    	{\frac{\bar{n}^2}{L^2} \left| s_{\sigma' \sigma_0}^{(1)}  s_{\sigma \sigma_0}^{(1)} \right|^2} 
  \nonumber \\
  &= 
    \left| 1 - \frac{4 \pi^2}{s_{\sigma' \sigma_0}^{(1)} s_{\sigma \sigma_0}^{(1)}} \sum_{l'_1 , l_1} e^{-i (\lambda_{l'_1}^{(1)} -\omega_0) \tau} \right.
  \nonumber \\
  & \qquad \qquad \times \left.
    \langle 0|   \tilde{b}_{\sigma} | 1, l'_1 \rangle \left\{  
      \frac{ 
        \langle \overline{1, l'_1} | \annihilate{\tilde{b}}{\sigma'} | 2 , l_2 \rangle 
        \langle \overline{2, l_2} | \create{\tilde{b}}{\sigma_0} | 1, l_1 \rangle  
      }{
        2 \omega_0 - \lambda_{l_2}^{(2)}
      }   
      - \frac{
        \langle \overline{1, l'_1} | \create{\tilde{b}}{\sigma_0} | 0 \rangle 
        \langle 0 |\annihilate{\tilde{b}}{\sigma'} | 1, l_1 \rangle
      }{
        \omega_0 - \lambda_{l'_1}^{(1)}
      } \right\} 
      \frac{\langle \overline{1, l_1} | \create{\tilde{b}}{\sigma_0} | 0 \rangle}{\omega_0 - \lambda_{l_1}^{(1)}} \right|^2 .
\end{align}
\end{widetext}
Note that the correlations are exponentially suppressed at large delay times $\tau$ with the decay rates $- \mathrm{Im} \lambda_{l'_1}^{(1)} $.

\bibliography{refs}

\end{document}